# EUNIS Habitat Maps: Enhancing Thematic and Spatial Resolution for Europe through Machine Learning

**Sara Si-Moussi[1], Stephan Hennekens[2], Sander Mücher[2], Wanda De Keersmaecker[3], Milan Chytrý[4], Emiliano Agrillo[5], Fabio Attorre[6], Idoia Biurrun[7], Gianmaria Bonari[8,9], Andraž Čarni[10,11], Renata Ćušterevska[12], Tetiana Dziuba[22], Klaus Ecker[13], Behlül Güler[14], Ute Jandt[15,16], Borja Jiménez-Alfaro[17,18], Jonathan Lenoir[19], Jens-Christian Svenning[20], Grzegorz Swacha[21], and Wilfried Thuiller[1*].**

1. Univ. Grenoble Alpes, Univ. Savoie Mont Blanc, CNRS, LECA, F-38000 Grenoble.
2. Wageningen Environmental Research (WENR), part of Wageningen University and Research (WUR), PO Box 47, 6700 AA, Wageningen, The Netherlands.
3. Flemish Institute for Technological Research (VITO), Mol, Belgium.
4. Department of Botany and Zoology, Faculty of Science, Masaryk University, Brno, Czech Republic.
5. Institute for Environmental Protection and Research (ISPRA), Rome, Italy
6. Sapienza University of Rome, Department of Environmental Biology, P.le Aldo Moro 5, 75 00185 Rome, Italy
7. Dept. Plant Biology and Ecology, University of the Basque Country UPV/EHU, Bilbao, Spain
8. Department of Life Sciences, University of Siena, Via Mattioli, 4, 53100, Siena, Italy
9. NBFC, National Biodiversity Future Center, 90133, Palermo, Italy
10. Research Centre of the Slovenian Academy of Sciences and Arts, Jovan Hadži Institute of Biology, Novi trg 2, 1000 Ljubljana, Slovenia
11. University of Nova Gorica, School for Viticulture and Enology, Vipavska cesta 13, 5000 Nova Gorica, Slovenia
12. Saints Cyril and Methodius University in Skopje Faculty of Natural Sciences and Mathematics: Skopje, Skopje, MK
13. Biodiversity and Conservation Biology Research, Swiss Federal Institute for Forest, Snow and Landscape Research WSL, Birmensdorf, Switzerland
14. Biology Education, Dokuz Eylul University, Buca, Izmir, Turkey
15. Martin Luther University Halle-Wittenberg, Institute of Biology / Geobotany and Botanical 20 Garden, Am Kirchtor 1, 06108 Halle, Germany
16. German Centre for Integrative Biodiversity Research (iDiv) Halle-Jena-Leipzig, Puschstrasse 22 4, Leipzig, 04103, Germany
17. Research Unit of Biodiversity (CSIC/UO/PA), University of Oviedo, Campus de Mieres, c/ 46 Gonzalo Gutiérrez Quirós s/n 33600 Mieres, Spain
18. Biodiversity Research Institute (IMIB), University of Oviedo–CSIC–Principality of Asturias, Mieres, Asturias, Spain
19. UMR CNRS 7058 "Ecologie et Dynamique des Systèmes Anthropisés" (EDYSAN), Université de Picardie Jules Verne, Amiens, France.
20. Center for Ecological Dynamics in a Novel Biosphere (ECONOVO), Department of Biology, Aarhus University, Ny Munkegade 114, DK-8000 Aarhus C, Denmark
21. Botanical Garden, University of Wrocław, Wrocław, Poland
22. Department of Geobotany and Ecology, M.G. Kholodny Institute of Botany, National Academy of Sciences of Ukraine, Kyiv, 2 Tereshchenkivska str. 01601, Kyiv, Ukraine

*Corresponding author(s): Wilfried Thuiller (wilfried.thuiller@univ-grenoble-alpes.fr) & Sara Si-Moussi (sara.si-moussi@univ-grenoble-alpes.fr)

## Abstract

The EUNIS habitat classification is crucial for categorising European habitats, supporting European policy on nature conservation and implementing the Nature Restoration Law. To meet the growing demand for detailed and accurate habitat information, we provide spatial predictions for 260 EUNIS habitat types at hierarchical level 3, together with independent validation and uncertainty analyses.

Using ensemble machine learning models, together with high-resolution satellite imagery and ecologically meaningful climatic, topographic and edaphic variables, we produced a European habitat map indicating the most probable EUNIS habitat at 100-m resolution across Europe. Additionally, we provide information on prediction uncertainty and the most probable habitats at level 3 within each EUNIS level 1 formation. This product is particularly useful for both conservation and restoration purposes.

Predictions were cross-validated at European scale using a spatial block cross-validation and evaluated against independent data from France (forests only), the Netherlands and Austria. The habitat maps obtained strong predictive performances on the validation datasets with distinct trade-offs in terms of recall and precision across habitat formations.

## Background & Summary

The European continent is very rich in natural and semi-natural habitats that host diverse species of flora and fauna, and provide a wide range of ecosystem services. However, these habitats are under major pressure due to climate change, pollution, biological invasions, rapid urbanisation, agricultural expansion and intensification in some areas and abandonment in others, which threaten the extent and quality of these habitats. The European Environmental Agency (EEA)'s latest assessment[1] on the State of Nature in Europe reveals an alarming decline in Europe's biodiversity, with most protected species and habitats lacking adequate conservation.

Despite these challenges, habitat assessments are largely based on expert judgment rather than field data[1], leading to uncertainties in evaluating their true condition. Additionally, the exact extent of habitats remains unknown, particularly outside the Natura 2000 network, making comprehensive conservation efforts more difficult. To monitor these pressing environmental pressures effectively, it is imperative to acquire accurate and comprehensive knowledge of the distribution of habitats at high spatial and thematic resolutions across Europe.

In Europe, the European Nature Information System (EUNIS)[2–4] is a habitat classification framework designed for a comprehensive coverage of habitat types across the continent. Unlike the Red List of Habitats[5], which focuses on assessing habitat risk status, EUNIS is particularly suited for large-scale mapping, including remote sensing-based "wall-to-wall" approaches. This also distinguishes it from widely used maps of Potential Natural Vegetation[6,7], which represent theoretical pre-human landscapes rather than current habitat distributions[8]. EUNIS consists of a hierarchical system with multiple nested levels, each offering increasing levels of detail and specificity in describing habitat types[4]. The classification system allows users to navigate from broader habitat categories to more specific habitat types (from level 1 to level 6). Level 1 distinguishes the major habitat formations as wetlands, grasslands, forests, etc. To align with the most recent revised classification, this study focuses on Level 3 which is the highest level with detailed descriptions for terrestrial habitats.

Producing habitat maps requires accurate in-situ data covering the diverse set of habitats in Europe[9]. The compilation of the European Vegetation Archive[10] (EVA) and advances in expert systems[4,11] have enabled large-scale classification of European habitats using in-situ vegetation data. These systems assign individual vegetation plots to established classification frameworks such as EUNIS, providing ground truth data for habitat modelling and mapping at large scale. However, while in-situ data provide valuable reference points, they are often spatially limited and time-consuming to collect.

Remote sensing offers a complementary approach by enabling large-scale, high-resolution habitat mapping across extensive and inaccessible areas. In addition to mapping habitat extent, remote sensing provides key environmental descriptors—such as vegetation indices (e.g., NDVI), surface moisture, canopy structure, and seasonal phenology— that can enhance habitat classification and potentially improve predictive models[12].

Recent studies showcase the potential of integrating remote sensing variables[13] with in-situ data for habitat modelling using knowledge-based classifiers[14], data-driven machine learning approaches[15,16] and hierarchical approaches[17]. However, a vast majority of studies focused on fine-scale mapping at regional scale, particularly within Natura 2000 networks[18,19]. Furthermore, most large-scale studies targeted a specific habitat group (e.g. Forests[20,21], coastal dunes[22], Grasslands[23]). While individual suitability maps for most EUNIS habitats at level 3 have been developed previously[24,25], integrating them into a single map proved challenging. Despite recent methodological advancements[26,27], a comprehensive, continental-scale map of EUNIS habitats has yet to be developed.

To support European conservation policies, there is a growing need to integrate different datasets and leverage the scalability and flexibility of Machine Learning (ML) methods for large-scale mapping of multiple habitats.

To this end, we developed and present high spatial and thematic resolution predictions of EUNIS (level 3) habitats across Europe. Here, we provide accurate habitat distribution maps at 100-m resolution by harnessing high-resolution and ecologically relevant remote sensing variables. We validated these novel habitat maps using three independent datasets and provided them to the community via public repositories.

# Methods

Habitat modelling predicts each habitat class at specific locations given environmental predictors. In this study, we used the EUNIS habitat nomenclature[4,28], focusing on terrestrial habitats at level 3. This includes over 250 distinct habitat classes within nine broader formations. To handle the discrete nature of habitat classes, we employed a classification approach to accurately assign habitat types to specific locations based on predictor variables available as gridded raster data at the European scale.

In practice, we built a set of multi-class machine learning models, where each model classifies data into one of multiple habitat classes. This contrasts with independent binary classifiers[24], which would require training separate models for each habitat class potentially leading to inconsistencies and loss of contextual relationships among classes. Joint modelling in multi-class models implicitly accounts for associations between multiple habitats, allowing less prevalent classes to borrow statistical strength from more common ones.

The presence of multiple habitats from different EUNIS level 1 formations within the same spatial unit can create mosaics, introducing ambiguity and potentially reducing model accuracy. To address this, we leveraged the hierarchical structure of habitat classes by training separate multi-class models at EUNIS level 3, each restricted to habitats within a single level 1 group, ensuring that each model focuses only on a subset of ecologically related habitats and aligning with the structure of the EUNIS system.

In the following, we detail the data sources (habitat in-situ information and environmental predictors), the detailed modelling and prediction strategies, the validation and finally the pipeline for producing spatially contiguous maps integrating all habitat classes i.e. *wall-to-wall* map.

## Data

### Study area

The study extent is defined by the EEA39 region. This includes the 27 EU member states, three EFTA countries (Iceland, Liechtenstein, and Norway), and nine additional collaborating countries: Albania, Bosnia and Herzegovina, Kosovo, Montenegro, North Macedonia, Serbia, Switzerland, Türkiye, and the United Kingdom. This extent aligns with the Corine land cover mask used further down for integrating habitat predictions into a wall-to-wall map.

## Habitat plots

Vegetation plots stored in the European Vegetation Archive (EVA)[10] served as ground truth data for training and testing the model across Europe. Each vegetation plot record was translated to the EUNIS typology at level 3 based on its species composition, using the EUNIS-ESy expert system[28]. The EUNIS classification includes nine revised habitat formations: MA (Marine habitats), N (Coastal habitats), P (Inland waters), Q (Wetlands), R (Grasslands), S (Shrublands - heathlands, scrub, and tundra), T (Forests), U (Sparsely vegetated habitats), and V (Man-made habitats).

An EVA vegetation plot typically contains a full list of co-occurring vascular plant species, often also a list of co-occurring bryophytes and lichens, estimates of cover-abundance of each species and various additional information on vegetation structure and layering. The dataset contained plots representing the target habitat groups (saltmarshes (MA2), coastal, wetlands, grasslands, shrublands, forests, sparsely vegetated and man-made habitats). Focusing on the terrestrial realm, we excluded other marine habitats and inland waters whose classification is not yet fully revised nor published.

Plots that were reported with no cover-abundance information for individual species were excluded. Further, plots smaller than 1 $m^2$, larger than 1000 $m^2$, without geographical coordinates and those with reported uncertainty of the coordinates larger than 100 m, were also excluded. The resulting dataset contained a total of about 597819 georeferenced plots, heterogeneously distributed across Europe (Figure 1, Table 1).

## Habitat datasets for validation

To evaluate the quality of the habitat maps, we have used two independent habitat occurrence datasets: a hold-out of habitat observations from the Netherlands (NLPT) and the French Forest Inventory (IFN). The NLPT dataset contains 50k vegetation plots (Figure 2) from the Landelijke Vegetatie Database (LVD)[29,30] sampled in the 2010-2022 period, covering diverse habitat classes. The French Forest Inventory, known as "Inventaire Forestier National"[31] (IFN), is a comprehensive program designed to assess and monitor, since 2005, the state of French forests at a yearly time step. The data is openly accessible through the DataIFN portal. The dataset contains about 180k habitat annotated forest plots (Figure 2).

Additionally, we conducted a regional validation using the Austrian (AT) habitat map[32] from 2013 (Figure 2) at high spatial resolution (10 m × 10 m grid, 8M grid cells) including all Austrian habitats (Figure 2).

### Environmental predictors

To train the models on the vegetation plots, we built a comprehensive database of environmental predictors, at the highest spatial resolution possible, which are ecologically meaningful to predict habitats across Europe[12,13]. Those variables needed to be available at least at 1 km resolution within our study area. We selected a set of the least correlated environmental variables including climate, topography, hydrography, geology, and soil (Table 2). These data were complemented by remote sensing (RS) products (Table 3) describing vegetation structure, phenology and productivity parameters and landscape composition.

## Ensemble machine learning framework

### Overview

To address uncertainties arising from model choices and data sampling, we employed an ensemble modelling approach[33,34].

First, the uncertainty arising from model choices stems from different algorithms with varying functional forms[35,36]. For instance, decision-tree approaches create trees with different depths, capturing interactions between variables whereas neural networks represent smooth, continuous responses, with wide architectures for recurring patterns and deep architectures for hierarchical representations[37,38]. To encompass the diversity of models, we created an ensemble of algorithms from different families of tree-based models as well as neural networks which meet a minimum performance requirement. Since decision trees excel with structured tabular data and neural networks with intricate feature interactions, combining them creates a more generalized and robust model, leveraging the strengths of each approach for improved predictive performance.

Second, we considered the uncertainty arising from different training data. Employing spatial block cross-validation[27,39], we trained the model with 20% of the observations hidden at each iteration. This process was repeated, generating an ensemble of classifiers with access to distinct samples, thus accounting for data sampling uncertainty.

### Selected ML algorithms

To achieve the best modelling performance, we employed well-known machine learning techniques, each with their advantages and disadvantages including bagging models, boosting models, and neural networks (Table 4).

Bagging models, also known as bootstrap aggregating models, enhance predictive accuracy and alleviate overfitting by training multiple individual estimator models on various subsets of both the training data and predictor variables. This approach harnesses collective knowledge to improve results. A notable example of bagging is the Random Forest (RF) algorithm[40], which employs classification or regression trees as base estimators.

Boosting models progressively train weak learners to create a robust learner by assigning greater emphasis to incorrectly classified instances. This iterative process leads to improved overall predictive performance.

- XGBoost[41]: This optimized gradient boosting algorithm merges tree-based models with regularization techniques, resulting in highly accurate and efficient predictions.
- CatBoost[42]: Tailored for categorical variable handling, CatBoost employs gradient-based strategies, ordered boosting, and innovative encoding methods to enhance accuracy and manage categorical features effectively.

  LightGBM[43]: A specialized form of boosting, LightGBM employs a gradient-based decision tree algorithm that optimizes training speed through leaf-wise growth and histogram-based optimizations, all while maintaining strong predictive performance.

Neural networks[44] are computational models that feature interconnected nodes, or "neurons" organised in layers. These networks learn to extract meaningful features by adjusting connection weights and biases during training. In this study, we used several architectures for fully Connected Neural Networks (Multi-layer Perceptron - MLP) with a single shallow hidden layer, a single wide hidden layer, two hidden layers and three hidden layers (Table 5).

## Dealing with habitat class imbalance

Class imbalance arises from the difference in prevalence amongst habitats due to a low of sampling effort or restricted extent in the case of threatened habitats. To address that, we explored class-imbalance correction approaches that do not involve data reduction or inflation. Instead, these methods modify the optimization objective to achieve class balance and as such are implemented differently depending on the machine learning algorithm.

For tree-based algorithms (RF, XGBoost, CatBoost and LightGBM), we evaluated class weighting, a technique that assigns to each class a weight inversely proportional to its frequency to balance its relative importance in the overall optimization.

For neural networks, the baseline loss function (optimization objective) for multi-class problems is the categorical cross-entropy. The categorical cross entropy[45] accepts a class weight parameter. When used, it is called weighted categorical cross-entropy (WCE).

Focal loss[46] (FL) is a modification of the standard Cross Entropy loss that specifically targets imbalanced classification problems. It introduces a focusing parameter "gamma" to down-weight the contribution of well-classified examples, putting more emphasis on hard, misclassified examples. This helps to alleviate the dominating effect of the majority class and enables the model to focus more on the minority class instances during model training.

Label Distribution Aware Margin[47] (LDAM) loss addresses class imbalance by assigning distinct margins to classes based on their distribution characteristics. These margins represent class boundary separation and control intra-class and inter-class distinctions. LDAM loss aims to penalize misclassifications of minority class examples more, encouraging the model to better account for underrepresented classes.

Focal Loss and LDAM can additionally incorporate a class-weighting parameter. Table 6 summarises the best imbalance strategy selected for each model for each habitat type at level 1.

## Ensemble model training

### Overview

For each habitat type at level 1, we trained an ensemble of multi-class models to predict the most likely EUNIS level 3 class within the target formation (i.e., level 1) following the steps depicted in Figure 3a:

1. Create a dataset containing observations from classes (EUNIS level 3 within each formation (EUNIS level 1).
2. Generate a spatial block partition of the selected observations for cross-validation.
3. Train an ensemble multi-class model for all classes within each formation, this step encompasses the feature pre-processing and hyperparameter tuning.
4. Select the best algorithm(s) in each family (bagging, boosting, neural networks) as well as the imbalance correction strategy based on overall cross-validation predictive performances.

In the following, we provide more details about the input data, pre-processing (feature pre-processing, data partitioning) and hyperparameter tuning steps.

## Input data

For training the models, the dataset consisted of the geolocated level 3 habitat observations from the EVA dataset and the abiotic and RS products predictors extracted at the highest spatial resolution available for the observations' locations. Similarly, for each external evaluation dataset (NLPT, AT IFN), the dataset consisted of the map of habitat observations annotated at level 3 and the abiotic and the RS predictor maps of the validation areas.

## Preprocessing steps

*Spatial Block-CV partitioning*

We establish a spatial block partition of the annotated dataset for cross-validation, as follows:

1. Grid Division: The study's spatial domain was divided into a grid composed of cells measuring 100 km × 100 km each;
2. Class Frequency Computation: Within each 100 km × 100 km cell, we calculated the frequencies of different habitat classes.
3. Cell Block Allocation: The 100 km × 100 km cells were then partitioned into five distinct spatial blocks, ensuring that every habitat class is adequately represented within each block. This is accomplished using the *IterativeStratification*[48] module found in the Python scikit-multilearn package[48];
4. Observation Assignment: Every individual habitat observation was assigned to the spatial block corresponding to the cell within which it resides.
5. Balanced Observations: We verified that the number of observations remained consistent across all blocks. If an imbalance was detected, we expanded the grid dimensions and initiated the process anew from step 1.

This spatial block partitioning methodology serves two crucial purposes. First, it amalgamates observations from proximate locations into the same partitions. This prevents potential overestimation of predictive performance stemming from data leakage induced by spatial autocorrelation. Second, in instances where multiple neighbouring habitats are observed and co-occurring in the same location, this technique exposes the model to a comprehensive array of potential responses for the same predictors. As a result, the model's probabilities reflect the uncertainty inherent in the dataset.

*Feature pre-processing*

We employed tailored feature pre-processing procedures customised to the nature of the features and the specific demands of the machine learning algorithms (Table 4). These pre-processing steps were seamlessly integrated within a unified pipeline that encompasses both algorithm training and prediction tasks.

## Hyperparameter tuning

For each individual machine learning algorithm, specific hyperparameters can be configured to control model complexity and optimization settings. To select the optimal hyperparameters for each modelling task (formation), a distinct 10% subset of the training data was set aside for hyperparameter tuning, with a focus on achieving high adjusted balanced accuracy as the objective. Our hyperparameter tuning process unfolds as follows:

1. We kept problem-specific parameters (such as objective function and metrics) at their default values, which are tailored for multi-class classification. These parameters are solely defined by the output variable type (here discrete classes).
2. Initially, a fixed set of architectures was generated, and we explored the best optimizer configurations without incorporating regularisation. For algorithms that are iteratively optimised, like boosting models and neural networks, we implemented early stopping callbacks to cease training when validation performance begins to decline, thereby avoiding overfitting.
3. Subsequently, we fine-tuned regularisation parameters to manage model complexity, leveraging a hold-out validation dataset. In the case of Multi-Layer Perceptrons (MLPs), given their computational complexity, we resorted to a grid search with a constrained array of configurations. We applied temperature scaling to MLPs to enhance the calibration of probabilities, especially when imbalance correction techniques are applied. This step was not required for bagging and boosting algorithms.
4. Bayesian optimization techniques were employed for bagging and boosting algorithms to navigate the hyperparameter space and identify the most optimal hyperparameter settings. This Bayesian hyperparameter tuning, facilitated by the *optuna* package[49], involves a probabilistic surrogate function known as the acquisition function. This function estimates the potential improvement in the objective function based on different hyperparameter combinations. The algorithm iteratively assesses various hyperparameter sets, updating the surrogate function and hyperparameter space accordingly.
5. Finally, we leveraged the performance outcomes on the test set to make informed selections of the best-performing algorithm(s) within each algorithm family.

This comprehensive and enhanced approach ensures the fine-tuning of algorithmic parameters, ultimately leading to selecting the most suitable models based on their predictive performance (Table 5).

Decision trees excel with structured tabular data and neural networks with intricate feature interactions. Therefore, combining them creates a more generalized and robust model, leveraging the strengths of each approach for improved predictive performance.

## Ensemble forecasting and uncertainty

Upon completion of the training process, we obtained a collection of trained machine learning algorithms alongside their associated feature pre-processing pipelines for every cross-validation fold. The ensemble model is a simple voting classifier, combining predictions from all individual models and assigning weights based on their respective rankings in terms of predictive performance (Figure 3b).

From the collective predictions of the ensemble, we extracted various metrics of uncertainty:

- *Model Uncertainty:* The variability in predictions among the constituent models within the ensemble indicates sensitivity to the choice of model.
- *Data Sampling Uncertainty:* The diversity in predictions across different folds within the ensemble underscores the sensitivity to variations in data sampling.

This uncertainty can be quantified at the level of individual pixels through different metrics, including:

- *Confidence Scores:* These scores represent the probability associated with the most probable class. Higher scores indicate more confidence in the chosen class, but the number of classes affects this measure. For example, a 30% confidence probability in a problem with 200 classes carries different implications than the same probability in a problem with just 3 classes.
- *Committee Averaging Scores:* These scores were calculated based on the proportion of voters (model x fold) that have predicted each class as the most likely. This measurement offers insights into the level of consensus or disagreement among models:
    - CA = 1: unanimous agreement among all models on prediction of the most likely habitat.
    - 0 < CA < 1: varying degrees of disagreement among models.

- CA ~ 0: Pronounced disagreement among models.

## Habitat mapping workflow

The ensemble multi-class habitat models combined with a set of decision rules were used to generate wall-to-wall habitat maps for Europe following these steps depicted in Figure 4.

### Step 1: Ensemble models

We used the ensemble model to predict the probabilities of each EUNIS level 3 class for each EUNIS 1 formation. Figure 5 shows an example map of habitat probability for the class of Fagus forests on non-acid soils (T17).

### Step 2: Regional filtering rules

Location is an intrinsic part of the EUNIS habitat classification. Some habitats are, by definition, associated with specific biogeographic regions (e.g., Macaronesian heathy forest). Although the climate can be a good approximator of biogeographic regions, we applied a post-hoc filtering to select the most likely habitat only amongst those which occur in the biogeographic region of the prediction pixel. This step also allowed to account the habitats' range of occurrence (e.g., Carpathian travertine fens).

For inland habitats, we used the ecoregions of the world[50]. For coastal habitats, we used the official EEA coastline delineation with an inland depth of 5 km away from the coastline. Using these layers and vegetation plot data from the EVA, we computed a matrix of association between each EUNIS level 3 class and the ecoregion/coastline, which were then used to generate regional masks for each modelled class.

**EUNIS class probabilities and ranking (output product n°1)** was generated by multiplying together the data cubes of the class-wise regional masks (step 2) and their model predicted probabilities (step 1). This output can be used to generate maps of the most likely habitat class at level 3 for each formation or aggregated at level 2. Figure 6 illustrates that for Scrub and Tundra (S) habitat classes, annotated at level 2 whereas Figures 7-8 illustrate that for broadleaved and coniferous forests respectively at level 3. These outputs are not filtered by land use and are therefore less sensitive to the minimum mapping unit of the land cover / land use maps.

### Step 3: Land cover filtering rules

Here, we used crosswalks to select for each EUNIS habitat class the associated land cover classes. From that, we generated land cover masks. The three most likely (top 3) **EUNIS classes and their confidence scores (output product n°2) within each EUNIS 1 formation** were obtained by multiplying together the data cubes of the class-wise land cover masks (step 3) with the EUNIS class probabilities (output product n°1). After rescaling, we selected the top 3 classes and their corresponding probabilities i.e. confidence scores. Only classes with non-zero probabilities were kept, therefore in some cases the top 2 and top 3 classes were undefined. Figure 9 shows an example of a map of the most likely habitat (top 1) for man-made vegetation habitats, accompanied by its confidence map.

### Step 4: Wall-to-wall mapping

Finally, we applied land cover-based priority rules (output product n°4) to determine the prevailing EUNIS 1 formation at each pixel to map the final habitat class at EUNIS level 3 (output product n°3).

Steps 1-2 were done only once at European scale for each EUNIS 1 formation.

Steps 3 and 4 required the definition of crosswalk rules tailored to the underlying land cover product. We provided the worksheet summarising the crosswalk rules for Corine Land Cover, which was used as a mask for producing the final habitat maps. The spatial resolution of the final layer as well as the habitat extents are thus controlled by the chosen land cover product. Any potential user can use its preferred land cover layer to refine the spatial resolution of the final product.

## Output products

All the following maps have been produced at 100 m resolution across Europe:

1. Continuous map of all EUNIS level 3 class probabilities.
2. Categorical map of the top 3 most likely habitats and their confidence scores at level 3 within each formation.
3. Categorical map of the top 3 EUNIS habitats at level 3 across all formations (previewed in Figure 10), with a legend and a QGIS style file.
4. An Excel sheet summarising the crosswalk and priority rules from Corine to EUNIS habitats.

**Technical specifications**:

- Spatial resolution: 100m
- Spatial extent: 900000.0000,7400000.0000,900000.0000,5500000.0000 [EPSG:3035]
- Time period: 2018-2022

## Data Records

The dataset is available in Zenodo under a Creative Commons Attribution 4.0 International, here: 10.5281/zenodo.11108226.

## Technical Validation

Habitat maps were cross validated at European scale on the EVA dataset and validated by comparison with independent datasets from the Netherlands (NLPT), Austria (AT) and France (IFN). We evaluated the predictive quality in terms of recall (*proportion of instances of a given habitat class correctly classified*), precision (*proportion of instances predicted as a given habitat class that truly belong to that class*), and F1-score (*harmonic mean of recall and precision*) for each EUNIS level 3 class. In multi-class settings, recall and precision often exhibit a trade-off because improving recall (capturing more true positives) can lead to an increase in false positives, lowering precision, while tightening classification to improve precision may exclude some true positives, reducing recall.

Table 7 summarizes the distribution of these class-wise metrics (mean and standard deviation) within each EUNIS level 1 formation. Detailed performances for EUNIS level 3 classes across the validation datasets, as well as cross-validation performances, are also provided respectively for Saltmarshes (Table 8), Coastal habitats (Table 9), Wetlands (Table 10), Grasslands (Table 11), Scrub and Tundra (Table 12), Forests (Table 13) and Sparsely vegetated habitats (Table 14).

The spatial-block cross-validation (EVA) results show strong predictive performance, with most classes achieving very good to acceptable F1-scores. Prediction quality varies depending on the number of habitat classes within the formation. Formations with fewer classes such as saltmarshes or those strongly shaped by abiotic elements (e.g. soil type, landscape structure) such as coastal, wetland, and sparsely vegetated habitats exhibit consistently high recall and precision. In contrast, grasslands, shrublands, and forests obtained variable predictive scores across classes reflecting their structural complexity and greater seasonal variability.

In sparsely vegetated habitats, precision and recall were well balanced. However, this was more the exception than the rule. Across formations, distinct trade-offs emerged: grasslands, shrublands, and forests exhibited higher precision, reflecting more conservative models, whereas coastal and wetland habitats showed higher recall, indicating a tendency toward over-prediction.

When evaluated on external datasets, the habitat maps performed well, albeit slightly lower than in cross-validation, except for coastal and salt marsh habitats. Similar trade-offs across formations persisted, with the largest performance drop observed for Austria, likely due to its higher resolution (10m) compared to the model's 100m resolution, making fine scale matching more challenging.

Finally, low performance was often associated with endemism prompting the use of an ecoregion filtering step or with habitats of limited extent, which improved when considering the top three predictions rather than only the most likely class.

## Known limitations and future improvements

### European habitat coverage

While we tried to be as exhaustive as possible in covering all habitat types in Europe, our map covers only a subset of EUNIS habitats represented in the EVA database which were classified by an expert system from plant community composition. Some of the missing habitats are challenging to define based on vegetation plot data (e.g., caves, glaciers) or in the case of anthropogenic habitats (e.g., tree plantations) hard to distinguish from semi-natural habitats. Habitats with a one-to-one association to a particular land cover class (e.g., glaciers, plantations) were incorporated the class in Step 4. Additionally, for statistical reasons, habitats with less than 5 occurrences in the curated EVA database were also discarded. Moreover, freshwater vegetation habitats were not mapped due to ongoing nomenclature revisions and lack of water quality predictors. Although there are some existing remote sensing products describing water body turbidity and trophic state, they are only available for a few large water bodies, and we found little overlap with freshwater vegetation plots. Finally, due to the choice of land use map (Corine land cover) which has a minimum mapping unit of 1ha, linear habitats (e.g. hedgerows, small streams) and other smaller extent habitats were sacrificed.

**Missing predictors**

Some habitat classes with low validation scores were linked to humid soil habitats (R53, R65), those affected by land use pressures such as grazing (R1N) and abandoned agricultural lands (S82). Future improvements should integrate predictors for soil moisture (beyond the included inundation occurrence), land use history, and human footprint. Additionally, certain low-performing classes, particularly those found at habitat edges such as woodland fringes (R53), could benefit from incorporating spatial landscape structure using deep convolutional neural networks with satellite imagery. The use of LiDAR imagery could also help in distinguishing classes with high structural complexity whereas incorporating seasonal remote sensing indicators could better capture the phenology of habitats.

**Validation scope**

We selected the validation areas based on available independent datasets covering different biogeographic regions: Alpine (Austria), Atlantic (Netherlands), Mediterranean, Alpine and Continental (IFN). There are many diverse regions that were not assessed due to a lack of independent data. Future efforts will be directed towards collecting more validation datasets to conduct this validation, particularly in the Mediterranean area, the Iberian Peninsula and Scandinavia.

# Code availability

The code to run the habitat modelling, evaluation and mapping workflow is provided in: https://github.com/bettasimousss/eunis-ml-mapping.git

## Acknowledgements

This work was carried out in the frame of the EO4DIVERSITY project funded by the European Space Agency through its Biodiversity+ Precursors programme. WT and S.SM also acknowledge funding from the Horizon Europe Natura Connect (No: 101060429) and OBSGESSION (No.: 101134954) projects. JCS was supported by Center for Ecological Dynamics in a Novel Biosphere (ECONOVO), funded by Danish National Research Foundation (grant DNRF173). G. Bonari was funded under the National Recovery and Resilience Plan (NRRP), Mission 4 Component 2 Investment 1.4 - Call for tender No. 3138 of 16 December 2021, rectified by Decree n. 3175 of 18 December 2021 of Italian Ministry of University and Research funded by the European Union – NextGenerationEU; Award Number: Project code CN_00000033, Concession Decree No. 1034 of 17 June 2022 adopted by the Italian Ministry of University and Research, CUP B63C22000650007, Project title "National Biodiversity Future Center - NBFC". We thank all data contributors to the European Vegetation Archive for providing vegetation plot data that supported our analysis.

## Author contributions

WT and S.SM designed the study in collaboration with SM and SH. S.SM designed the framework and ran all the models and predictions. SH provided the EVA, NLPT and Austrian datasets. WT and SM provided the financial support. WT wrote the initial draft of the manuscript with the help of all co-authors.

## Competing interests

The authors declare no competing interest.

# Tables

Table 1. *Number of vegetation plots for each EUNIS level 1 habitat formation*

| Saltmarshes (MA2) | Coastal (N) | Wetlands (Q) | Grassland (R) | Shrubland (S) | Forest (T) | Sparsely vegetated (U) | Man-made (V) |
|---|---|---|---|---|---|---|---|
| 6552 | 20287 | 57764 | 217523 | 39343 | 186844 | 5893 | 53161 |

Table 2. *List of environmental predictors used for habitat modeling and mapping at 100m resolution.*

| **Type** | **Predictor** | **Data source** |
|---|---|---|
| **Climate** | BIO1: Annual mean temperature (°C) | CHELSA V2.1[1]<br>Resolution: 1km<br>Temporal range: 1981-2010<br>Spatial range:<br>Europe |
| | BIO4 : Temperature seasonality – bio 4 | |
| | GDD5: Growing degree days heat sum above 5°C (°C) | |
| | BIO12: Annual sum of precipitationq (kg.m²/yr) | |
| | BIO15: Precipitation seasonality | |
| | SCD: Snow covered days (day count) | |
| | SWE: Snow water equivalent (kg.m²/yr) | |
| **Topography** | EU DEM slope (degrees) | EU DEM v1.1[2]<br>Resolution: 100m |
| | EU DEM aspect (degrees) | |
| | Landform classification based on the topographic position index.<br>(10 classes) | EU-DEM v.1.1.[2], upscaled to 100m, on SAGA-GIS.<br>Resolution: 100m |
| **Hydrography** | Distance to inland water (m) | EU-Hydro river network[3]<br>Resolution: 100m |
| | Distance to sea coastline (m) | EU-Hydro coast-line[3]<br><br>Resolution: 100m |
| **Geology** | Dominant parent material class<br><br>(27 classes) | European soil database (ESDB)[4]<br><br>Resolution: 1km |
| | Depth to rock (m) | |
| **Soil** | Available Water Capacity (AWC) for the topsoil fine earth fraction | LUCAS topsoil physical properties [5]<br><br>Resolution: 500m |
| | Bulk density of the topsoil (kg/dm$^3$) | |
| | Coarse fragment (%) content in topsoil | |
| | Sand, silt, and clay proportions (%) | |
| | Soil acidity | LUCAS topsoil chemical properties[6]<br><br>Resolution: 500m |
| | Organic carbon content (g/kg) | |
| | Nitrogen content (g/kg) | |
| | Calcium carbonates (g/kg) | |
| | Cation exchange capacity (cmol/kg) | |

*Table 3. Remote sensing products used for habitat modeling and mapping at 100m resolution.*

| Type | Predictor | Data source |
|---|---|---|
| **Vegetation phenology and productivity** | Season amplitude given by MAXV-MINV | Copernicus Land Monitoring Service (CLMS), VITO [7]<br><br>Resolution: 10m |
| | Length of season (number of days between start and end) | |
| | Slope of the green-up season (*PPI* × *day*-1) | |
| | PPI at the day of maximum-of-season | |
| | Total productivity (PPI × day) | |
| **Leaf area Index** | Leaf area index in summer ($m^2/m^2$) | Copernicus Land Monitoring Service (CLMS) [8] |
| | Leaf area index in spring ($m^2/m^2$) | |
| **Inundation** | Inundation seasonality | High-resolution mapping of global surface water and its long-term changes [9]<br><br>Resolution: 100m<br><br>Temporal range: 1984-2021 |
| **Canopy structure** | Tree canopy cover density (%) | Copernicus Land Monitoring Service (CLMS) [8]<br><br>Resolution: 10m |
| | Height of the tree canopy (m) | Lang et al 2023 [10]l<br><br>Resolution: 10m |
| **Land cover** | Proportion (%) of pixels of each landcover class in a 100m radius | Source: ESA WorldCover [11]<br><br>Resolution: 100m<br><br>Temporal coverage: 2020 |

*Table 4. Specific feature preprocessing routines per feature type and algorithm family.*

| Feature type | Bagging algorithms | Boosting algorithms | Neural networks |
|---|---|---|---|
| **Continuous**<br><br>(e.g., temperature, pH) | / | / | Center-scale |
| **Ordinal**<br><br>(e.g., inundation occurrence) | MinMax scaling | | |
| **Cyclical / periodic**<br><br>(e.g., aspect, SOSD) | Cyclical transformer: transforms the value into an angle on a circular space, then applies a mapping to the cosinus-sinus coordinates.<br><br>Example:<br><br>- DEM aspect<br>- Cosinus(aspect): east/west slope orientation<br>- Sinus(aspect): north/south slope orientation | | |
| **Categorical**<br><br>(e.g., TPI_landform) | One-hot-encoding | Ordinal encoding | Categorical embedding |
| **Day counts**<br><br>(e.g., season LENGTH) | Scaling by 365 | | |
| **Frequency**<br><br>(e.g., landscape composition) | Scaling by the total frequency (e.g., 100) | | |

*Table 5. Common algorithm configurations including shared hyperparameters.*

| Algorithm family | Algorithm (implementation) | Configuration |
|---|---|---|
| Bagging | Random Forest (scikit-learn[12]) | • Criterion: gini<br>• Number of trees: 128<br>• Maximum features: sqrt(n_features)<br>• Maximum samples: 1.0 |
| Boosting | XGBoost (xgboost[13]) | • Objective: multi:softprob<br>• Booster: gbtree<br>• Tree method: histogram<br>• Maximum boosting iterations: 512<br>• Early stopping rounds: 10<br>• Min_delta: 0.001 |
| | LightGBM (lightgbm[14]) | • Objective: softmax<br>• Booster: gbdt<br>• Maximum boosting iterations: 512<br>• Early stopping rounds: 10<br>• Min_delta: 0.001 |
| | CatBoost (catboost[15]) | • Objective: MultiClass<br>• Iterations: 100 |
| Neural networks | MLP (pytorch-lightning[16])<br><br>Architectures:<br>• MLP1_shallow (MLPs): [128]<br>• MLP1_wide (MLPw): [1024]<br>• MLP2: [256,128]<br>• MLP3: [512,256,128] | • Batch norm: True<br>• Batch size: 1024<br>• Maximum epochs: 100<br>• Early stopping rounds: 10 |

*Table 6. Selected models per algorithm family for each formation with optimal hyperparameters and class imbalance correction.*

| **EUNIS formation** | **Algorithm family** | **Algorithm** | **Configuration** |
|---|---|---|---|
| MA2, N, P, Q | Bagging | Random Forest | **Class-weighted Random Forest**<br>● Maximum depth: 6<br>● Class weight: balanced |
| | Boosting | XGBoost | **Class-weighted XGBoost**<br>● Learning rate: 0.1<br>● Maximum depth: 5<br>● Sample weights: True |
| | Neural networks | MLP3 | **MLP3 with weighted LDAM loss**<br>● Architecture: MLP3<br>● Loss: LDAM, class weights, margin=0.5<br>● Optimizer: RAdam (lr=0.001, decay=0.001) |
| R | Bagging | Random Forest | **Random Forest**<br>● Maximum depth: 11<br>● Class weight: uniform |
| | Boosting | XGBoost | **XGBoost**<br>● Learning rate: 0.1<br>● Maximum depth: 6<br>● Sample weight: False |
| | Neural networks | MLPw | **MLPw with unweighted LDAM loss**<br>● Loss: LDAM, uniform weights, margin=0.5<br>● Optimizer: RAdam (lr=0.001, decay=0.001) |
| | | | **MLP3 with unweighted LDAM loss**<br>● Max epochs: 100<br>● Loss: LDAM, uniform weights, margin=0.5<br>● Optimizer: RAdam (lr=0.001, decay=0.001) |
| S | Bagging | Random Forest | **Random Forest**<br>● Maximum depth: 11<br>● Class weight: uniform |
| | Boosting | XGBoost | **XGBoost**<br>● Learning rate: 0.1<br>● Maximum depth: 6 |
| | | LightGBM | **LightGBM** |
| | | | • Learning rate: 0.01<br>• Maximum depth: 6<br>• Class weight: None |
| | Neural networks | MLP | **MLPw with unweighted LDAM loss**<br>• Loss: LDAM, margin=0.5<br>• Optimizer: RAdam (lr=0.001, decay=0.001) |
| | | | **MLP3 with Focal loss**<br>• Loss: Focal loss (gamma=5.0)<br>• Optimizer: SGD (momentum=0.9, nesterov=True, lr=0.001, decay=0.001) |
| T | Bagging | Random Forest | **Class-weighted Random Forest**<br>• Maximum depth: 6<br>• Class weight: balanced |
| | Boosting | XGBoost | **XGBoost**<br>• Learning rate: 0.1<br>• Maximum depth: 6 |
| | | LightGBM | **LightGBM**<br>• Learning rate: 0.01<br>• Maximum depth: 6<br>• Class weight: None |
| | | CatBoost | **Class-weighted CatBoost**<br>• Auto_class_weights: Balanced<br>• Learning rate: 0.2 |
| | Neural networks | MLP | **MLPw with Focal Loss**<br>• Loss: Focal loss (gamma = 5.0)<br>• Optimizer: SGD (momentum=0.9, nesterov=True, lr=0.001, decay=0.001) |
| | | | **MLP3 with Focal Loss**<br>• Loss: Focal loss (gamma = 5.0)<br>• Optimizer: SGD (momentum=0.9, nesterov=True, lr=0.001, decay=0.001) |
| U | Bagging | RandomForest | **Random Forest**<br>• Maximum depth: 6 |
| | Boosting | LightGBM | **LightGBM**<br>• Learning rate: 0.01<br>• Maximum depth: 6 |

<table>
<tr><td rowspan="5"></td><td rowspan="2"></td><td></td><td><ul><li>Class weight: None</li></ul></td></tr>
<tr><td>XGBoost</td><td>XGB<ul><li>Learning rate: 0.1</li><li>Maximum depth: 6</li></ul></td></tr>
<tr><td rowspan="2">Neural networks</td><td rowspan="2">MLP</td><td>MLPs with LDAM loss<ul><li>Architecture: [128]</li><li>Max_epochs: 100</li><li>Loss: LDAM (margin=0.5)</li><li>Optimizer: RAdam (lr=0.001, decay=0.001)</li></ul></td></tr>
<tr><td>MLPw with LDAM loss<ul><li>Loss: LDAM (margin = 0.5)</li><li>Optimizer: RAdam (lr=0.001, decay=0.001)</li></ul></td></tr>
</table>

Table 7 - F1-score, precision and recall from the ensemble forecasting models for the EVA and the three validation datasets: Netherlands (NLPT), Austria (AU) and the French Forest Inventory (IFN).

| Strategy | F1-score | | | | Precision | | | | Recall | | | |
|---|---|---|---|---|---|---|---|---|---|---|---|---|
| | EVA | NLPT | AU | IFN | EVA | NLPT | AU | IFN | EVA | NLPT | AU | IFN |
| **MA2 - Littoral biogenic habitats** | 0.87 ± 0.11 | 0.76 ± 0.34 | | | 0.82 ± 0.18 | 0.74 ± 0.33 | | | 0.95 ± 0.06 | 0.80 ± 0.36 | | |
| **N – coastal habitats** | 0.81 ± 0.09 | 0.82 ± 0.09 | | | 0.75 ± 0.15 | 0.84 ± 0.06 | | | 0.92 ± 0.07 | 0.83 ± 0.18 | | |
| **Q – Wetlands** | 0.73 ± 0.13 | 0.95 ± 0.06 | 0.56 ± 0.46 | | 0.67 ± 0.19 | 0.95 ± 0.08 | 0.55 ± 0.45 | | 0.86 ± 0.09 | 0.96 ± 0.07 | 0.57 ± 0.46 | |
| **R – Grasslands** | 0.66 ± 0.17 | 0.40 ± 0.32 | 0.37 ± 0.35 | | 0.78 ± 0.15 | 0.57 ± 0.42 | 0.45 ± 0.39 | | 0.59 ± 0.18 | 0.37 ± 0.33 | 0.37 ± 0.36 | |
| **S - Scrub and Tundra** | 0.83 ± 0.12 | 0.46 ± 0.39 | 0.31 ± 0.42 | | 0.89 ± 0.06 | 0.49 ± 0.41 | 0.31 ± 0.42 | | 0.79 ± 0.15 | 0.46 ± 0.41 | 0.31 ± 0.43 | |
| **T – Forests** | 0.61 ± 0.20 | 0.38 ± 0.31 | 0.33 ± 0.39 | 0.48 ± 0.32 | 0.77 ± 0.15 | 0.47 ± 0.36 | 0.34 ± 0.40 | 0.78 ± 0.32 | 0.55 ± 0.23 | 0.37 ± 0.35 | 0.32 ± 0.39 | 0.40 ± 0.33 |
| **U - Sparsely vegetated** | 0.94 ± 0.03 | 0.81 ± 0.22 | 0.35 ± 0.40 | | 0.95 ± 0.03 | 0.96 ± 0.05 | 0.34 ± 0.39 | | 0.94 ± 0.03 | 0.78 ± 0.31 | 0.36 ± 0.23 | |

Table 8 - *Predictive performances of littoral biogenic habitat classes (MA2) at level 3*

| Class | EVA | | | NLPT | | |
|---|---|---|---|---|---|---|
| | **F1-score** | **Precision** | **Recall** | **F1-score** | **Precision** | **Recall** |
| **MA211** | 0.922 | 0.855 | 1.000 | - | - | - |
| **MA221** | 0.651 | 0.483 | 1.000 | 0.906 | 0.828 | 1.000 |
| **MA222** | 0.834 | 0.730 | 0.971 | 0.934 | 0.881 | 0.993 |
| **MA223** | 0.882 | 0.935 | 0.835 | 0.929 | 0.913 | 0.945 |
| **MA224** | 0.882 | 0.917 | 0.850 | 0.900 | 0.953 | 0.852 |
| **MA225** | 0.672 | 0.514 | 0.972 | 0.912 | 0.838 | 1.000 |
| **MA232** | 0.992 | 0.984 | 1.000 | - | - | - |
| **MA241** | 0.999 | 0.999 | 0.999 | - | - | - |
| **MA251** | 0.852 | 0.751 | 0.983 | - | - | - |
| **MA252** | 0.922 | 0.904 | 0.941 | - | - | - |
| **MA253** | 0.947 | 0.984 | 0.912 | - | - | - |

Table 9 - *Predictive performances of coastal habitat classes (N) at level 3*

| Class | **EVA** | | | **NLPT** | | |
|---|---|---|---|---|---|---|
| | **F1-score** | **Precision** | **Recall** | **F1-score** | **Precision** | **Recall** |
| **N11** | 0.699 | 0.549 | 0.961 | 0.850 | 0.855 | 0.844 |
| **N12** | 0.774 | 0.688 | 0.885 | - | - | - |
| **N13** | 0.871 | 0.927 | 0.821 | 0.828 | 0.721 | 0.973 |
| **N14** | 0.826 | 0.914 | 0.754 | - | - | - |
| **N15** | 0.848 | 0.875 | 0.823 | 0.629 | 0.940 | 0.473 |
| **N16** | 0.835 | 0.897 | 0.781 | - | - | - |
| **N17** | 0.871 | 0.809 | 0.944 | - | - | - |
| **N18** | 0.647 | 0.489 | 0.957 | 0.732 | 0.882 | 0.625 |
| **N19** | 0.703 | 0.554 | 0.962 | 0.894 | 0.808 | 1.000 |
| **N1A** | 0.857 | 0.890 | 0.826 | 0.755 | 0.898 | 0.650 |
| **N1B** | 0.815 | 0.712 | 0.952 | - | - | - |
| **N1C** | 0.916 | 0.845 | 1.000 | - | - | - |
| **N1D** | 0.817 | 0.731 | 0.928 | 0.875 | 0.792 | 0.977 |
| **N1F** | 0.809 | 0.685 | 0.986 | - | - | - |
| **N1G** | 0.849 | 0.743 | 0.991 | - | - | - |
| **N1H** | 0.878 | 0.932 | 0.829 | 0.888 | 0.810 | 0.981 |
| **N1J** | 0.821 | 0.699 | 0.993 | - | - | - |
| **N21** | 0.780 | 0.689 | 0.898 | 0.928 | 0.921 | 0.935 |
| **N22** | 0.547 | 0.376 | 1.000 | - | - | - |
| **N31** | 0.929 | 0.915 | 0.945 | 0.824 | 0.808 | 0.840 |
| **N32** | 0.920 | 0.905 | 0.935 | - | - | - |
| **N33** | 0.887 | 0.819 | 0.966 | - | - | - |
| **N34** | 0.826 | 0.703 | 1.000 | - | - | - |
| **N35** | 0.701 | 0.543 | 0.988 | - | - | - |

Table 10 - *Predictive performances for wetland classes (Q) at level 3*

| Class | EVA | | | NLPT | | | AU | | |
|---|---|---|---|---|---|---|---|---|---|
| | **F1-score** | **Precision** | **Recall** | **F1-score** | **Precision** | **Recall** | **F1-score** | **Precision** | **Recall** |
| **Q11** | 0.769 | 0.764 | 0.774 | 0.744 | 0.842 | 0.667 | 0.830 | 0.842 | 0.818 |
| **Q12** | 0.770 | 0.642 | 0.963 | - | - | - | 0.885 | 0.794 | 1.000 |
| **Q21** | 0.754 | 0.648 | 0.903 | 0.782 | 0.690 | 0.903 | - | - | - |
| **Q22** | 0.754 | 0.754 | 0.755 | 0.713 | 0.868 | 0.605 | 0.779 | 0.815 | 0.746 |
| **Q23** | 0.489 | 0.325 | 0.990 | - | - | - | - | - | - |
| **Q24** | 0.757 | 0.748 | 0.766 | 0.460 | 0.890 | 0.310 | - | - | - |
| **Q25** | 0.739 | 0.762 | 0.717 | 0.606 | 0.795 | 0.490 | 0.699 | 0.865 | 0.586 |
| **Q41** | 0.779 | 0.906 | 0.684 | 0.500 | 1.000 | 0.333 | 0.872 | 0.825 | 0.925 |
| **Q42** | 0.701 | 0.601 | 0.841 | 0.483 | 0.778 | 0.350 | - | - | - |
| **Q43** | 0.653 | 0.513 | 0.897 | 0.667 | 0.500 | 1.000 | - | - | - |
| **Q44** | 0.580 | 0.421 | 0.931 | - | - | - | - | - | - |
| **Q45** | 0.803 | 0.719 | 0.910 | - | - | - | 0.796 | 0.833 | 0.762 |
| **Q46** | 0.261 | 0.150 | 1.000 | - | - | - | - | - | - |
| **Q51** | 0.827 | 0.974 | 0.718 | 0.851 | 0.764 | 0.959 | - | - | - |
| **Q52** | 0.810 | 0.849 | 0.774 | 0.780 | 0.742 | 0.822 | - | - | - |
| **Q53** | 0.767 | 0.713 | 0.829 | 0.872 | 0.876 | 0.868 | - | - | - |
| **Q54** | 0.688 | 0.541 | 0.943 | 1.000 | 1.000 | 1.000 | - | - | - |
| **Q61** | 0.751 | 0.650 | 0.889 | 0.576 | 0.865 | 0.432 | - | - | - |
| **Q62** | 0.778 | 0.670 | 0.928 | 0.917 | 0.888 | 0.947 | - | - | - |
| **Q63** | 0.670 | 0.506 | 0.992 | 1.000 | 1.000 | 1.000 | - | - | - |

Table 11 - *Predictive performances for grassland classes (R) at level 3*

| Class | EVA | | | NLPT | | | AU | | |
|---|---|---|---|---|---|---|---|---|---|
| | F1-score | Precision | Recall | F1-score | Precision | Recall | F1-score | Precision | Recall |
| **R11** | 0.832 | 0.890 | 0.781 | - | - | - | 0.231 | 0.796 | 0.135 |
| **R12** | 0.775 | 0.954 | 0.653 | - | - | - | - | - | - |
| **R13** | 0.682 | 0.831 | 0.579 | 0.522 | 1.000 | 0.353 | - | - | - |
| **R14** | 0.795 | 0.868 | 0.733 | - | - | - | - | - | - |
| **R15** | 0.539 | 0.813 | 0.404 | - | - | - | - | - | - |
| **R16** | 0.724 | 0.784 | 0.673 | - | - | - | - | - | - |
| **R17** | 0.513 | 0.850 | 0.367 | - | - | - | - | - | - |
| **R18** | 0.854 | 0.908 | 0.807 | - | - | - | - | - | - |
| **R19** | 0.786 | 0.925 | 0.683 | - | - | - | - | - | - |
| **R1A** | 0.806 | 0.784 | 0.828 | 0.364 | 0.251 | 0.667 | 0.213 | 0.130 | 0.586 |
| **R1B** | 0.721 | 0.738 | 0.704 | - | - | - | - | - | - |
| **R1C** | 0.687 | 0.740 | 0.640 | - | - | - | - | - | - |
| **R1D** | 0.711 | 0.847 | 0.612 | - | - | - | - | - | - |
| **R1E** | 0.833 | 0.872 | 0.797 | - | - | - | - | - | - |
| **R1F** | 0.752 | 0.852 | 0.673 | - | - | - | - | - | - |
| **R1G** | 0.654 | 0.702 | 0.613 | - | - | - | - | - | - |
| **R1H** | 0.633 | 0.693 | 0.582 | - | - | - | - | - | - |
| **R1J** | 0.757 | 0.858 | 0.677 | - | - | - | - | - | - |
| **R1K** | 0.692 | 0.830 | 0.594 | - | - | - | - | - | - |
| **R1M** | 0.693 | 0.861 | 0.580 | 0.559 | 0.904 | 0.405 | 0.309 | 0.858 | 0.189 |
| **R1N** | 0.431 | 0.835 | 0.290 | - | - | - | - | - | - |
| **R1P** | 0.641 | 0.758 | 0.554 | 0.610 | 0.934 | 0.453 | 0.697 | 0.835 | 0.599 |
| **R1Q** | 0.576 | 0.746 | 0.469 | 0.870 | 0.843 | 0.898 | - | - | - |
| **R1R** | 0.832 | 0.888 | 0.783 | - | - | - | - | - | - |
| **R1S** | 0.363 | 0.606 | 0.259 | - | - | - | 0.773 | 0.772 | 0.774 |
| **R21** | 0.708 | 0.705 | 0.712 | 0.658 | 0.733 | 0.597 | 0.649 | 0.861 | 0.521 |
| **R22** | 0.777 | 0.698 | 0.878 | 0.778 | 0.641 | 0.989 | 0.866 | 0.820 | 0.918 |
| **R23** | 0.776 | 0.848 | 0.715 | - | - | - | 0.684 | 0.748 | 0.630 |
| **R24** | 0.469 | 0.605 | 0.383 | - | - | - | - | - | - |
| **R31** | 0.701 | 0.739 | 0.667 | - | - | - | - | - | - |
| **R32** | 0.435 | 0.833 | 0.294 | - | - | - | - | - | - |
| **R33** | 0.496 | 0.505 | 0.488 | - | - | - | - | - | - |
| **R34** | 0.726 | 0.805 | 0.661 | - | - | - | - | - | - |
| **R35** | 0.692 | 0.696 | 0.689 | 0.397 | 0.938 | 0.251 | 0.541 | 0.640 | 0.469 |
| **R36** | 0.697 | 0.784 | 0.628 | 0.845 | 0.847 | 0.844 | - | - | - |
| **R37** | 0.679 | 0.790 | 0.596 | 0.236 | 0.912 | 0.136 | 0.646 | 0.786 | 0.549 |
| **R41** | 0.707 | 0.895 | 0.584 | - | - | - | 0.623 | 0.489 | 0.858 |
| **R42** | 0.539 | 0.495 | 0.592 | - | - | - | 0.824 | 0.973 | 0.715 |
| **R43** | 0.798 | 0.815 | 0.782 | - | - | - | - | - | - |
| **R44** | 0.793 | 0.777 | 0.810 | - | - | - | 0.943 | 0.899 | 0.990 |
| **R45** | 0.805 | 0.849 | 0.765 | - | - | - | - | - | - |
| **R51** | 0.617 | 0.925 | 0.462 | - | - | - | 0.222 | 0.719 | 0.131 |
| **R52** | 0.783 | 0.944 | 0.669 | - | - | - | - | - | - |
| **R53** | 0.230 | 1.000 | 0.130 | - | - | - | - | - | - |
| **R54** | 0.757 | 0.893 | 0.656 | 0.405 | 1.000 | 0.254 | - | - | - |
| **R55** | 0.740 | 0.814 | 0.678 | 0.773 | 0.927 | 0.663 | 0.645 | 0.813 | 0.535 |
| **R56** | 0.754 | 0.829 | 0.691 | - | - | - | 0.844 | 0.841 | 0.847 |
| **R57** | 0.756 | 0.934 | 0.634 | 0.649 | 0.900 | 0.507 | - | - | - |
| **R61** | 0.611 | 0.799 | 0.494 | - | - | - | - | - | - |
| **R62** | 0.636 | 0.708 | 0.578 | - | - | - | 0.761 | 0.669 | 0.882 |
| **R63** | 0.673 | 0.829 | 0.566 | - | - | - | - | - | - |
| **R64** | 0.522 | 0.694 | 0.418 | - | - | - | - | - | - |
| **R65** | 0.251 | 0.707 | 0.153 | - | - | - | - | - | - |

Table 12 - *Predictive performances for scrub and tundra classes (S) at level 3*

| Class | EVA | | | NLPT | | | AU | | |
|---|---|---|---|---|---|---|---|---|---|
| | **F1-score** | **Precision** | **Recall** | **F1-score** | **Precision** | **Recall** | **F1-score** | **Precision** | **Recall** |
| **S11** | 0.816 | 0.801 | 0.831 | - | - | - | - | - | - |
| **S12** | 0.646 | 0.840 | 0.525 | - | - | - | - | - | - |
| **S21** | 0.875 | 0.931 | 0.826 | - | - | - | - | - | - |
| **S22** | 0.892 | 0.842 | 0.948 | - | - | - | 0.980 | 0.991 | 0.969 |
| **S23** | 0.849 | 0.876 | 0.823 | - | - | - | - | - | - |
| **S24** | 0.849 | 0.977 | 0.751 | - | - | - | - | - | - |
| **S25** | 0.873 | 0.895 | 0.853 | - | - | - | 0.865 | 0.799 | 0.944 |
| **S26** | 0.908 | 0.885 | 0.931 | - | - | - | 0.892 | 0.857 | 0.930 |
| **S31** | 0.875 | 0.928 | 0.827 | 0.790 | 0.870 | 0.723 | 0.773 | 0.940 | 0.656 |
| **S32** | 0.886 | 0.922 | 0.852 | 0.472 | 0.935 | 0.315 | - | - | - |
| **S33** | 0.873 | 0.906 | 0.844 | 0.790 | 0.821 | 0.762 | - | - | - |
| **S34** | 0.985 | 0.998 | 0.973 | - | - | - | - | - | - |
| **S35** | 0.895 | 0.898 | 0.891 | 0.819 | 0.708 | 0.972 | - | - | - |
| **S36** | 0.866 | 0.906 | 0.830 | - | - | - | - | - | - |
| **S37** | 0.882 | 0.931 | 0.838 | - | - | - | - | - | - |
| **S38** | 0.876 | 0.935 | 0.824 | 0.862 | 0.922 | 0.810 | - | - | - |
| **S41** | 0.881 | 0.902 | 0.860 | 0.882 | 0.815 | 0.961 | - | - | - |
| **S42** | 0.909 | 0.881 | 0.938 | 0.765 | 0.644 | 0.940 | 0.920 | 0.887 | 0.955 |
| **S43** | 0.772 | 0.880 | 0.688 | - | - | - | - | - | - |
| **S51** | 0.899 | 0.857 | 0.945 | - | - | - | - | - | - |
| **S52** | 0.876 | 0.953 | 0.810 | - | - | - | - | - | - |
| **S53** | 0.888 | 0.976 | 0.815 | - | - | - | - | - | - |
| **S54** | 0.909 | 0.914 | 0.904 | - | - | - | - | - | - |
| **S61** | 0.889 | 0.853 | 0.927 | - | - | - | - | - | - |
| **S62** | 0.758 | 0.950 | 0.631 | - | - | - | - | - | - |
| **S63** | 0.837 | 0.859 | 0.816 | - | - | - | - | - | - |
| **S64** | 0.669 | 0.748 | 0.606 | - | - | - | - | - | - |
| **S65** | 0.829 | 0.940 | 0.741 | - | - | - | - | - | - |
| **S66** | 0.762 | 0.883 | 0.670 | - | - | - | - | - | - |
| **S67** | 0.868 | 0.850 | 0.886 | - | - | - | - | - | - |
| **S68** | 0.736 | 0.787 | 0.691 | - | - | - | - | - | - |
| **S71** | 0.853 | 0.947 | 0.776 | - | - | - | - | - | - |
| **S72** | 0.800 | 0.822 | 0.779 | - | - | - | - | - | - |
| **S73** | 0.626 | 0.815 | 0.508 | - | - | - | - | - | - |
| **S74** | 0.859 | 0.836 | 0.883 | - | - | - | - | - | - |
| **S75** | 0.859 | 0.871 | 0.847 | - | - | - | - | - | - |
| **S76** | 0.796 | 0.845 | 0.752 | - | - | - | - | - | - |
| **S81** | 0.816 | 0.810 | 0.823 | - | - | - | - | - | - |
| **S82** | 0.211 | 1.000 | 0.118 | - | - | - | - | - | - |
| **S91** | 0.892 | 0.905 | 0.880 | 0.814 | 0.835 | 0.794 | 0.949 | 0.960 | 0.938 |
| **S92** | 0.880 | 0.877 | 0.884 | 0.748 | 0.852 | 0.666 | 0.816 | 0.761 | 0.880 |
| **S93** | 0.862 | 0.877 | 0.848 | - | - | - | - | - | - |
| **S94** | 0.810 | 0.873 | 0.755 | - | - | - | - | - | - |

Table 13 - *Predictive performances for forest habitat classes (T) at level 3*

| Class | EVA | | | NLPT | | | AU | | | IFN | | |
|---|---|---|---|---|---|---|---|---|---|---|---|---|
| | F1 | Precision | Recall | F1 | Precision | Recall | F1 | Precision | Recall | F1 | Precision | Recall |
| T11 | 0.753 | 0.847 | 0.677 | 0.551 | 0.944 | 0.390 | 0.720 | 0.825 | 0.638 | 0.439 | 0.935 | 0.287 |
| T12 | 0.715 | 0.747 | 0.685 | 0.440 | 0.897 | 0.292 | 0.702 | 0.774 | 0.642 | 0.173 | 0.865 | 0.096 |
| T13 | 0.738 | 0.811 | 0.677 | 0.711 | 0.714 | 0.708 | - | - | - | 0.479 | 0.911 | 0.325 |
| T14 | 0.625 | 0.702 | 0.563 | - | - | - | - | - | - | 0.444 | 1.000 | 0.286 |
| T15 | 0.650 | 0.739 | 0.581 | 0.817 | 0.720 | 0.944 | 0.717 | 0.747 | 0.690 | 0.154 | 0.714 | 0.086 |
| T16 | 0.667 | 0.809 | 0.567 | 0.380 | 0.904 | 0.240 | 0.831 | 0.933 | 0.749 | 0.235 | 1.000 | 0.133 |
| T17 | 0.784 | 0.724 | 0.855 | 0.692 | 0.624 | 0.777 | 0.890 | 0.865 | 0.917 | 0.901 | 0.830 | 0.986 |
| T18 | 0.729 | 0.801 | 0.670 | 0.678 | 0.838 | 0.569 | - | - | - | 0.758 | 0.884 | 0.663 |
| T19 | 0.793 | 0.758 | 0.831 | - | - | - | 0.441 | 0.388 | 0.510 | 0.895 | 0.923 | 0.870 |
| T1A | 0.576 | 0.863 | 0.432 | - | - | - | - | - | - | - | - | - |
| T1B | 0.773 | 0.810 | 0.738 | 0.526 | 0.681 | 0.429 | 0.622 | 0.857 | 0.488 | 0.323 | 0.818 | 0.201 |
| T1C | 0.568 | 0.838 | 0.430 | 0.693 | 0.558 | 0.912 | - | - | - | - | - | - |
| T1D | 0.436 | 0.892 | 0.289 | 0.091 | 0.500 | 0.050 | - | - | - | - | - | - |
| T1E | 0.764 | 0.708 | 0.830 | - | - | - | 0.804 | 0.776 | 0.833 | 0.716 | 0.829 | 0.631 |
| T1F | 0.735 | 0.809 | 0.674 | - | - | - | - | - | - | 0.320 | 0.884 | 0.196 |
| T1G | 0.491 | 0.696 | 0.380 | - | - | - | 0.887 | 0.824 | 0.961 | - | - | - |
| T1H | 0.817 | 0.916 | 0.737 | 0.634 | 0.659 | 0.612 | 0.836 | 0.930 | 0.759 | - | - | - |
| T21 | 0.764 | 0.738 | 0.792 | - | - | - | - | - | - | 0.850 | 1.000 | 0.739 |
| T22 | 0.503 | 0.750 | 0.378 | - | - | - | - | - | - | - | - | - |
| T23 | 0.236 | 0.760 | 0.140 | - | - | - | - | - | - | - | - | - |
| T24 | 0.706 | 0.824 | 0.617 | - | - | - | - | - | - | - | - | - |
| T25 | 0.404 | 0.875 | 0.263 | - | - | - | - | - | - | - | - | - |
| T27 | 0.567 | 0.842 | 0.427 | 0.609 | 0.750 | 0.512 | - | - | - | - | - | - |
| T28 | 0.107 | 0.375 | 0.062 | - | - | - | - | - | - | - | - | - |

| | | | | | | | | | | | | | |
|---|---|---|---|---|---|---|---|---|---|---|---|---|---|
| **T29** | 0.370 | 0.818 | 0.239 | - | - | - | - | - | - | - | - | - |
| **T31** | 0.770 | 0.756 | 0.784 | - | - | - | 0.897 | 0.904 | 0.891 | 0.614 | 0.956 | 0.453 |
| **T32** | 0.764 | 0.825 | 0.711 | - | - | - | - | - | - | 0.888 | 0.930 | 0.849 |
| **T33** | 0.538 | 0.767 | 0.414 | - | - | - | - | - | - | - | - | - |
| **T34** | 0.761 | 0.838 | 0.696 | - | - | - | 0.949 | 0.955 | 0.943 | 0.896 | 0.832 | 0.971 |
| **T35** | 0.712 | 0.669 | 0.761 | - | - | - | 0.839 | 0.856 | 0.822 | 0.693 | 0.972 | 0.538 |
| **T36** | 0.698 | 0.833 | 0.601 | - | - | - | - | - | - | - | - | - |
| **T37** | 0.640 | 0.848 | 0.514 | - | - | - | - | - | - | - | - | - |
| **T38** | 0.547 | 0.665 | 0.465 | - | - | - | - | - | - | - | - | - |
| **T39** | 0.743 | 0.931 | 0.618 | - | - | - | - | - | - | - | - | - |
| **T3A** | 0.807 | 0.834 | 0.782 | - | - | - | - | - | - | - | - | - |
| **T3B** | 0.258 | 0.645 | 0.161 | - | - | - | - | - | - | - | - | - |
| **T3C** | 0.558 | 0.893 | 0.406 | - | - | - | - | - | - | - | - | - |
| **T3D** | 0.619 | 0.733 | 0.535 | - | - | - | - | - | - | - | - | - |
| **T3E** | 0.082 | 1.000 | 0.043 | - | - | - | - | - | - | - | - | - |
| **T3F** | 0.812 | 0.794 | 0.831 | - | - | - | - | - | - | - | - | - |
| **T3G** | 0.835 | 0.783 | 0.894 | - | - | - | - | - | - | - | - | - |
| **T3H** | 0.541 | 0.780 | 0.414 | - | - | - | - | - | - | - | - | - |
| **T3J** | 0.701 | 0.798 | 0.625 | - | - | - | - | - | - | - | - | - |
| **T3K** | 0.670 | 0.838 | 0.558 | - | - | - | - | - | - | 0.235 | 1.000 | 0.133 |
| **T3M** | 0.774 | 0.762 | 0.787 | 0.727 | 0.586 | 0.957 | - | - | - | - | - | - |

Table 14 - *Predictive performances for sparsely vegetated habitat classes (U) at level 3*

| Class | EVA | | | NLPT | | | AU | | |
|---|---|---|---|---|---|---|---|---|---|
| | **F1-score** | **Precision** | **Recall** | **F1-score** | **Precision** | **Recall** | **F1-score** | **Precision** | **Recall** |
| **U21** | 0.916 | 0.887 | 0.947 | - | - | - | - | - | - |
| **U22** | 0.939 | 0.910 | 0.971 | 1.000 | 1.000 | 1.000 | 0.859 | 0.872 | 0.847 |
| **U24** | 0.943 | 0.941 | 0.945 | - | - | - | - | - | - |
| **U25** | 0.943 | 0.957 | 0.930 | - | - | - | - | - | - |
| **U26** | 0.909 | 0.926 | 0.893 | - | - | - | 0.862 | 0.853 | 0.870 |
| **U27** | 0.964 | 0.967 | 0.960 | - | - | - | 0.871 | 0.810 | 0.942 |
| **U28** | 0.946 | 0.976 | 0.917 | - | - | - | - | - | - |
| **U29** | 0.910 | 0.899 | 0.921 | - | - | - | - | - | - |
| **U31** | - | - | - | - | - | - | 0.827 | 0.852 | 0.804 |
| **U32** | 0.919 | 0.940 | 0.898 | - | - | - | - | - | - |
| **U33** | 0.961 | 0.973 | 0.949 | 0.500 | 1.000 | 0.333 | - | - | - |
| **U34** | 0.962 | 0.985 | 0.940 | - | - | - | - | - | - |
| **U35** | 0.865 | 0.889 | 0.842 | - | - | - | 0.857 | 0.840 | 0.874 |
| **U36** | 0.924 | 0.939 | 0.909 | - | - | - | - | - | - |
| **U37** | 0.965 | 0.962 | 0.968 | 0.941 | 0.889 | 1.000 | - | - | - |
| **U38** | 0.952 | 0.948 | 0.956 | - | - | - | - | - | - |
| **U3B** | 0.966 | 0.977 | 0.956 | - | - | - | - | - | - |
| **U3D** | 0.984 | 0.995 | 0.973 | - | - | - | - | - | - |
| **U3E** | - | - | - | - | - | - | 0.827 | 0.852 | 0.804 |
| **U42** | - | - | - | - | - | - | 0.783 | 0.660 | 0.962 |
| **U62** | 0.990 | 1.000 | 0.981 | - | - | - | - | - | - |
| **U71** | 0.973 | 0.971 | 0.974 | - | - | - | - | - | - |
| **U72** | 0.960 | 1.000 | 0.923 | - | - | - | - | - | - |

# Figures

**Figure 1 – Distribution and density (log-scaled, 100km x 100km grid) of vegetation plots from the European Vegetation Archive (EVA) used in this study**

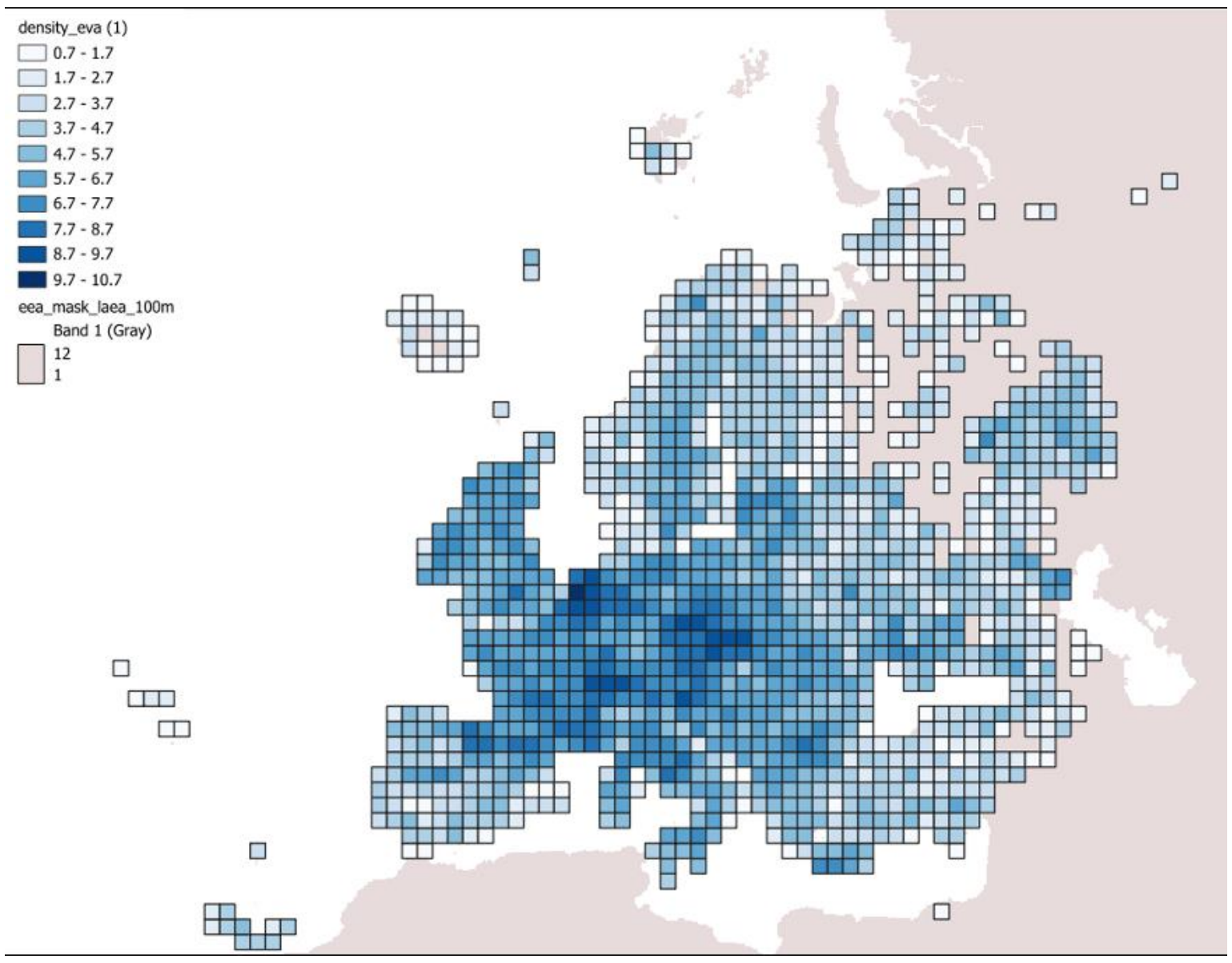

**Figure 2 - Distribution of habitat observations used for validation and number of observations in each EUNIS formation.**

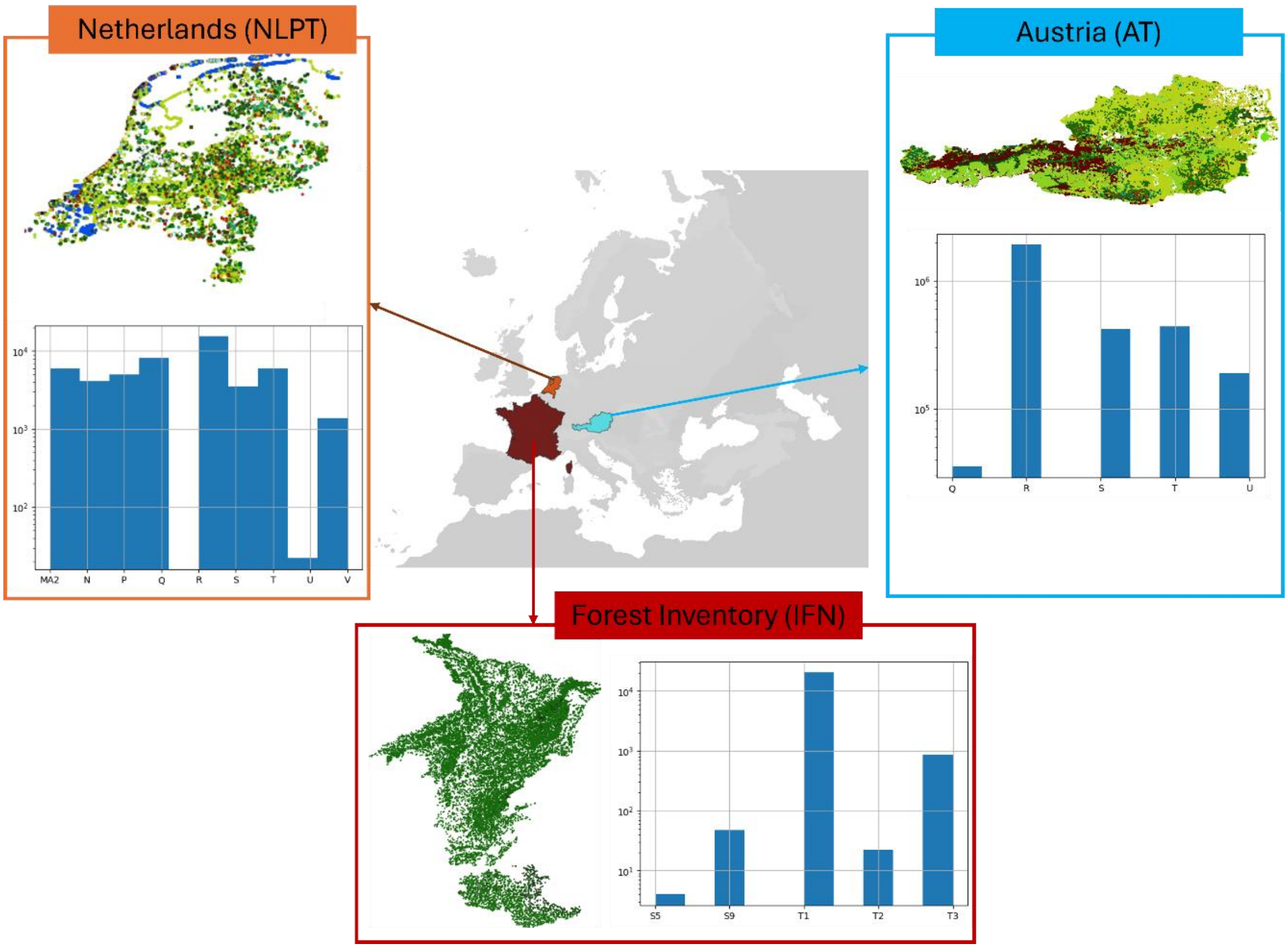

**Figure 3 - Ensemble multi-class modeling framework**

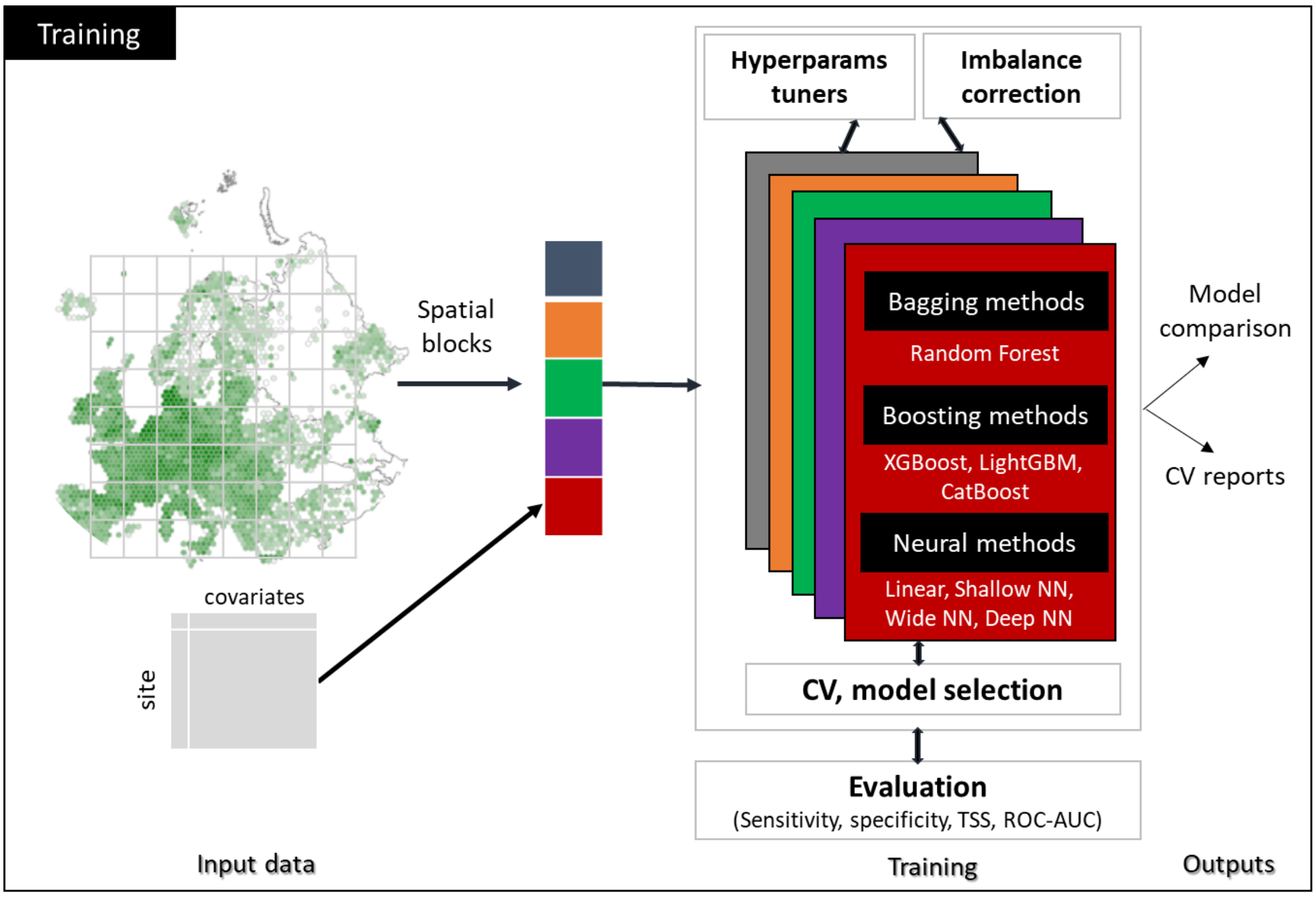

(a) Ensemble model training

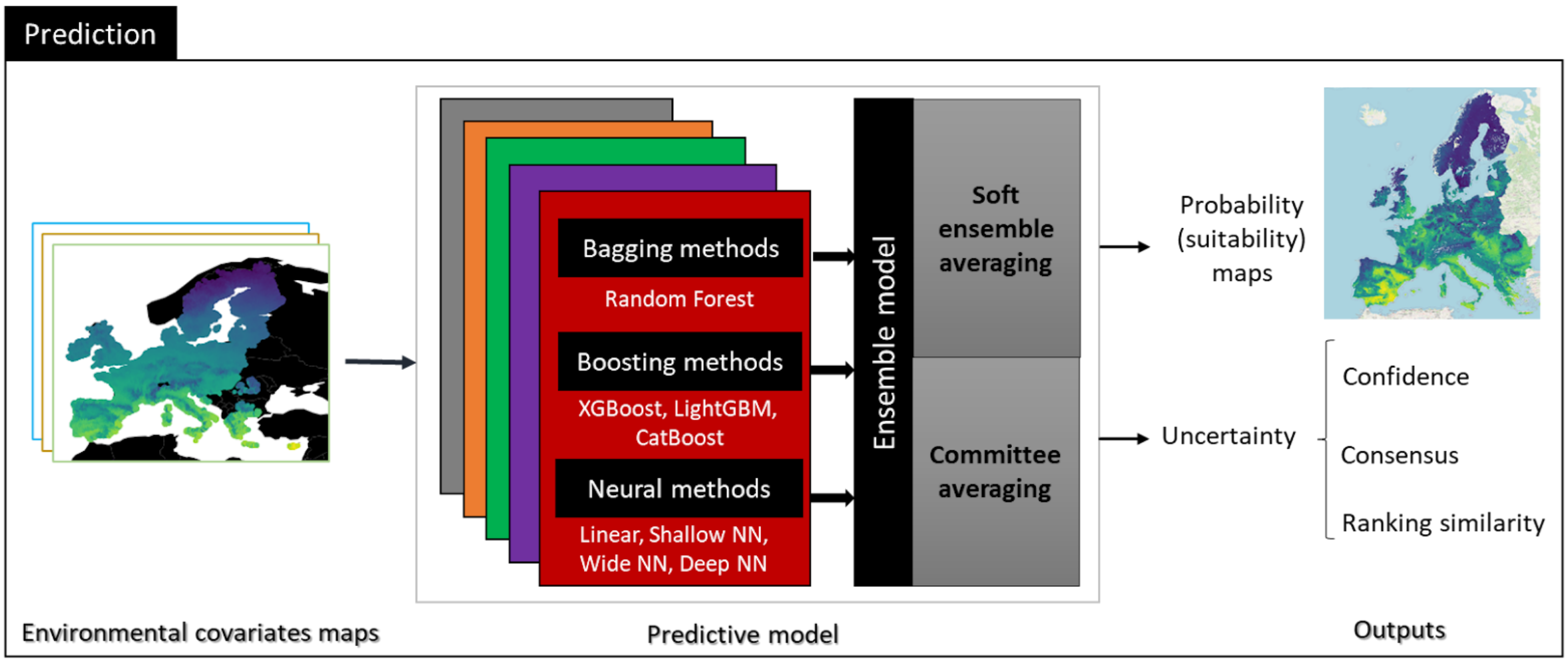

(b) Ensemble forecasting and uncertainty

## Figure 4 - Workflow to produce European habitat maps

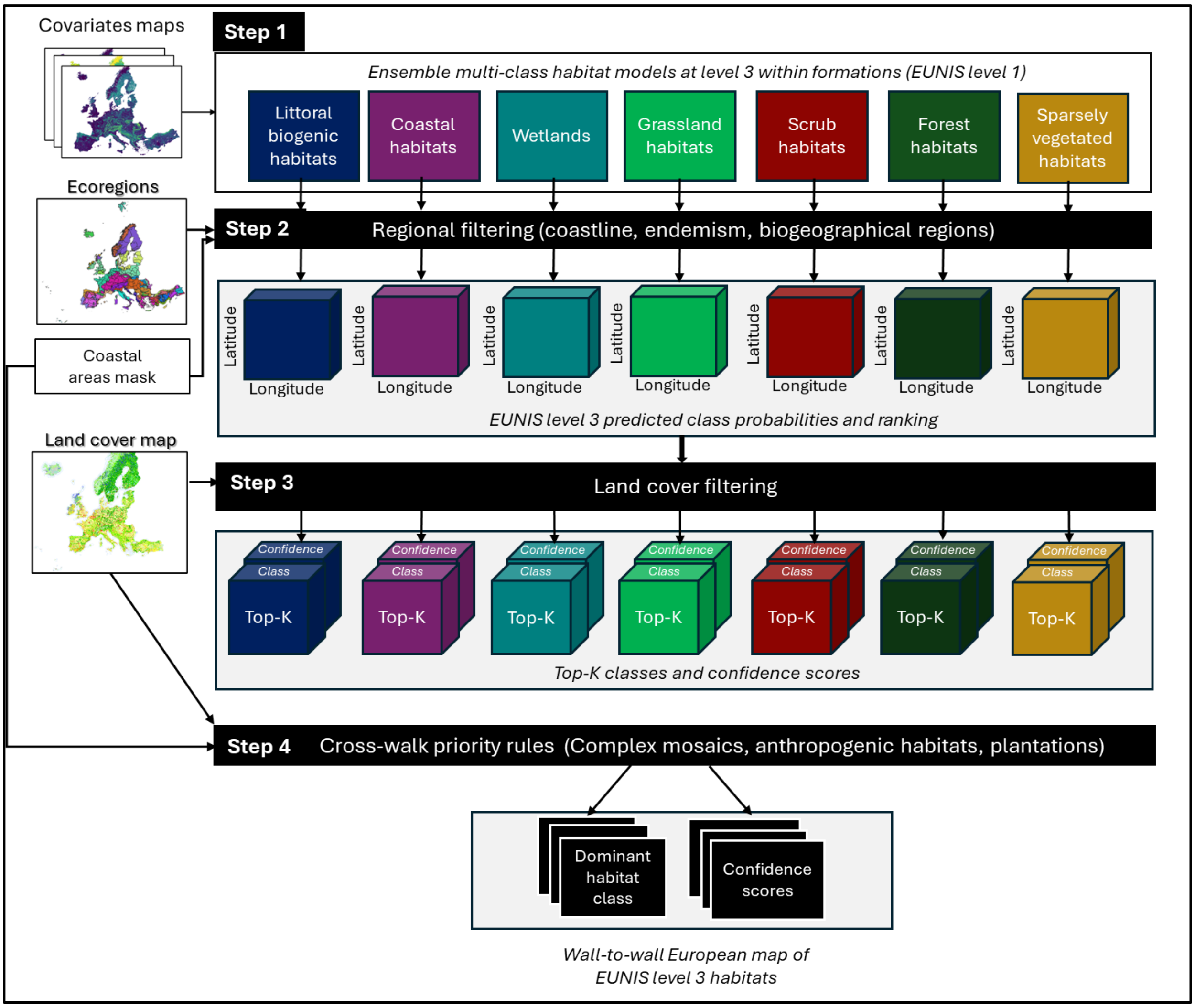

**Figure 5 – Habitat probability map of T17 « Fagus forest on non-acid soils ».**

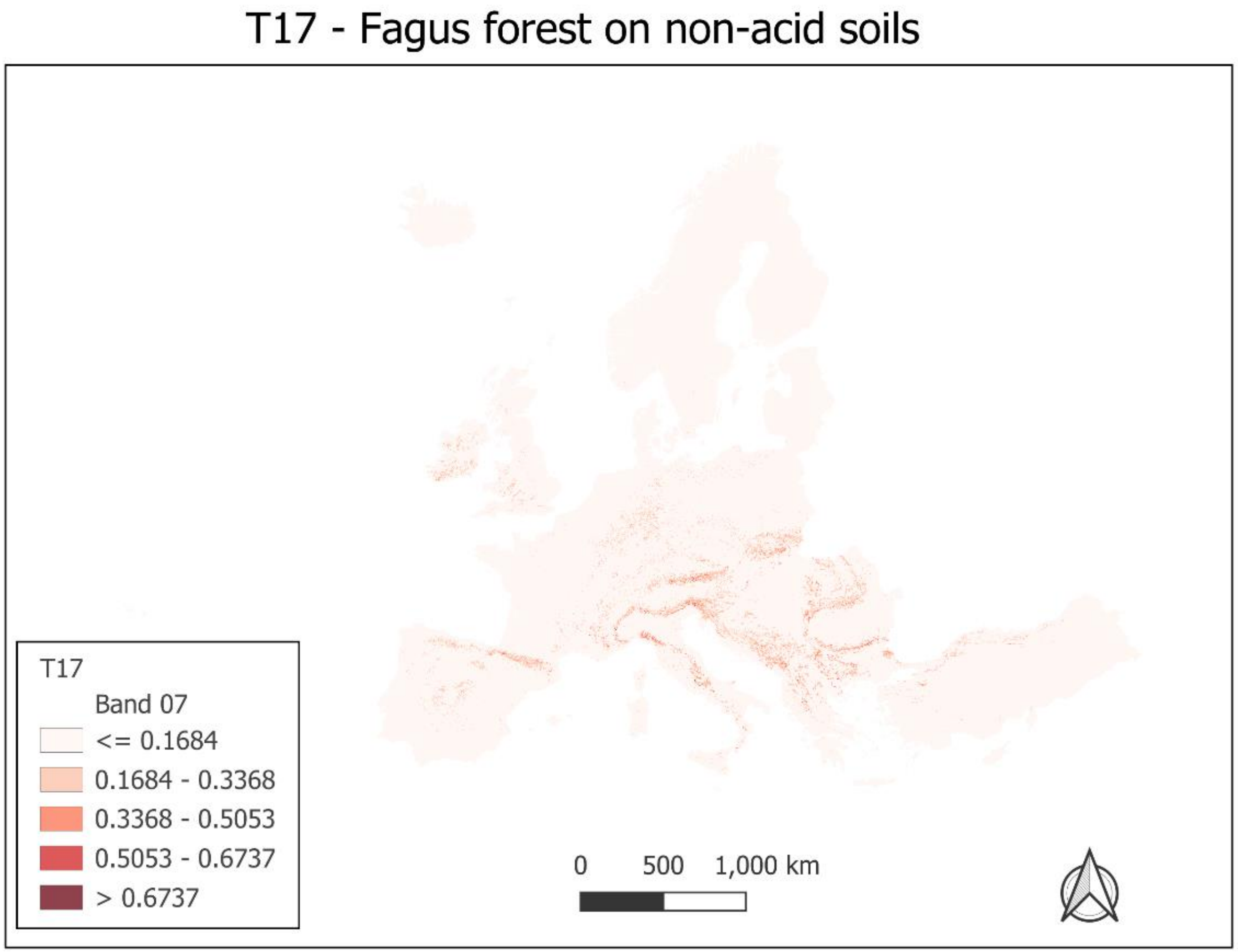

**Figure 6 – Dominant scrub and tundra habitat map with color legend at EUNIS level 2**

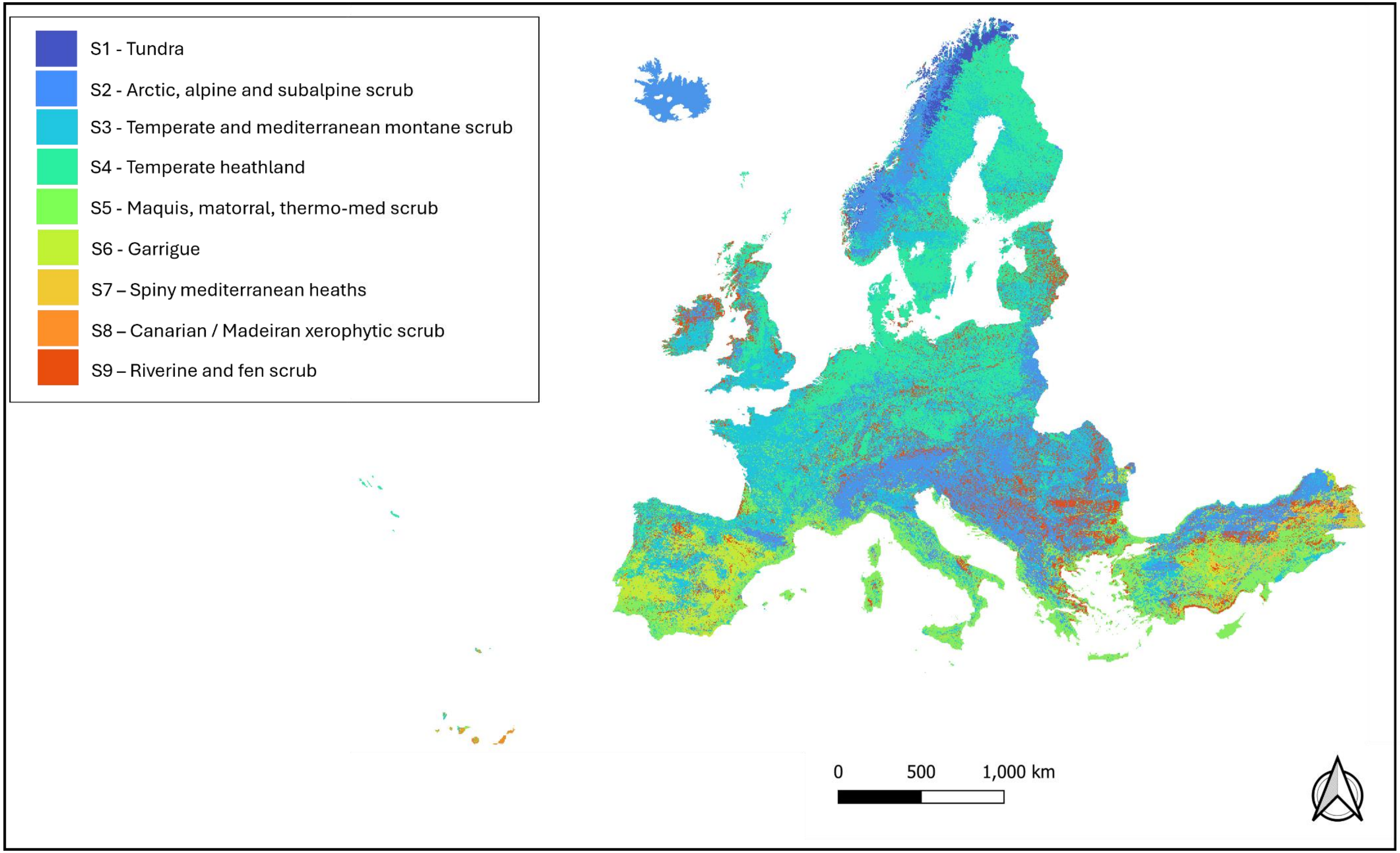

**Figure 7 – Dominant broadleaved forest habitat map at EUNIS level 3**

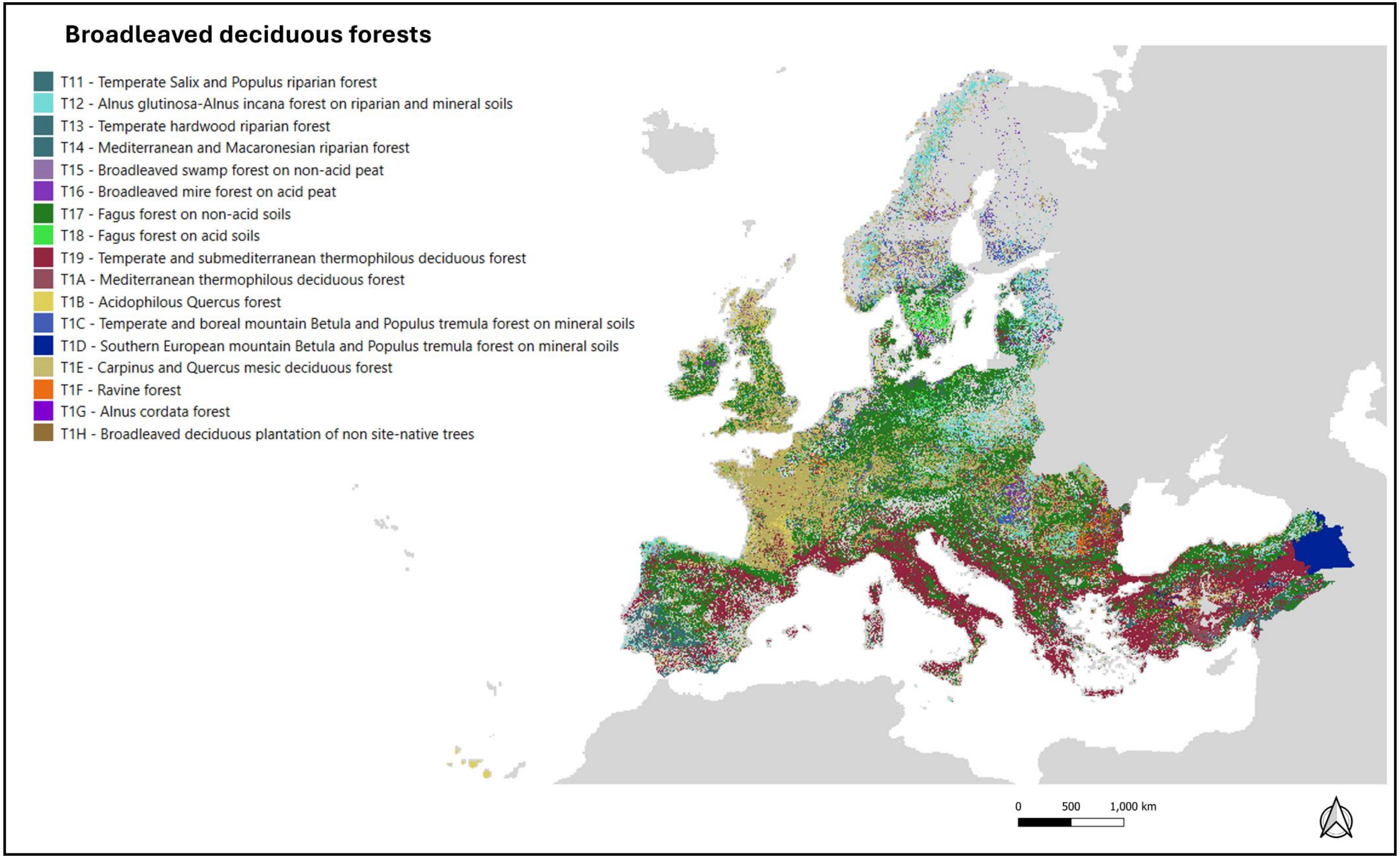

**Figure 8 – Dominant coniferous forests habitat map at EUNIS level 3**

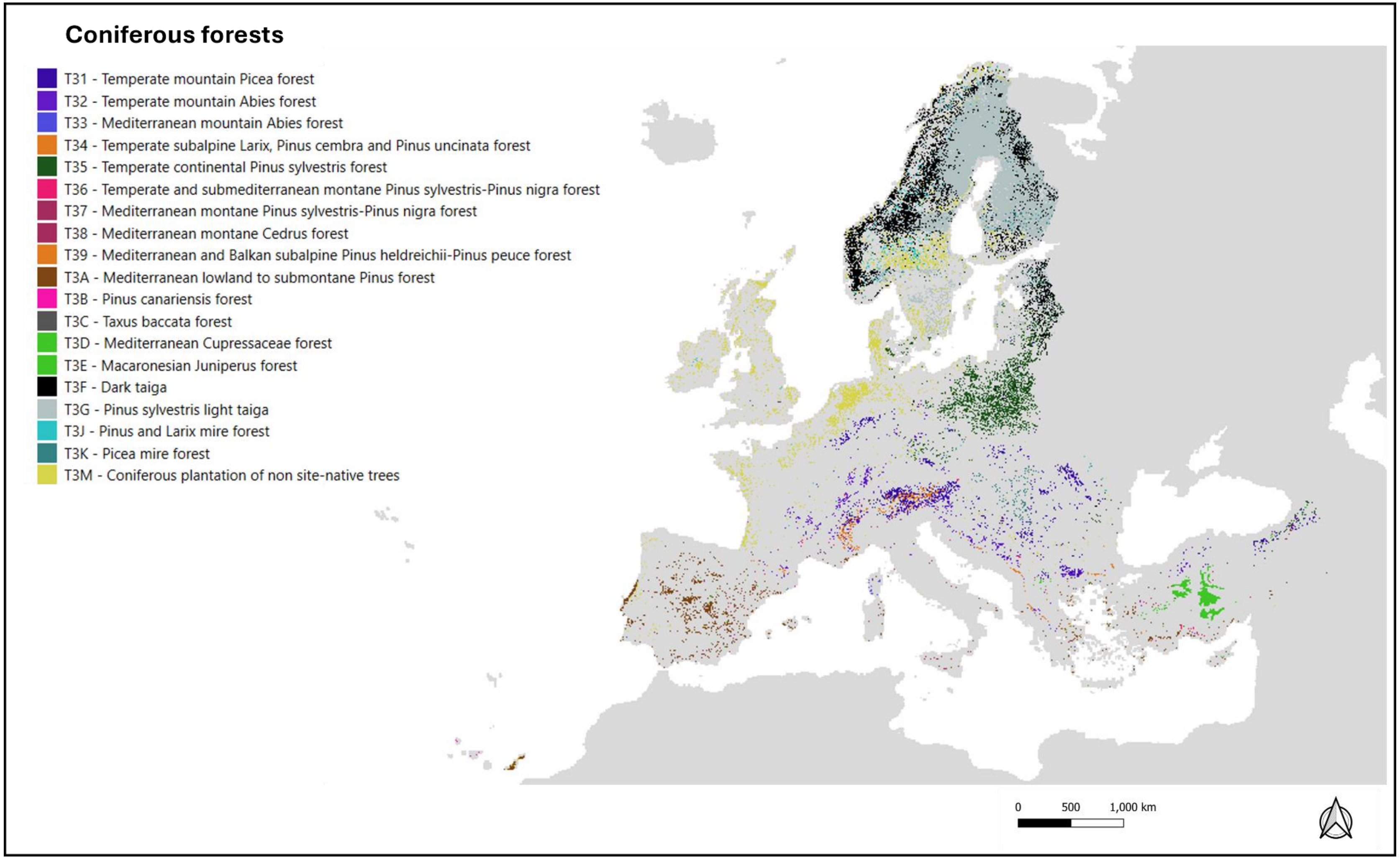

**Figure 9 – Dominant man-made vegetation habitat class (EUNIS level 3) and confidence map.**

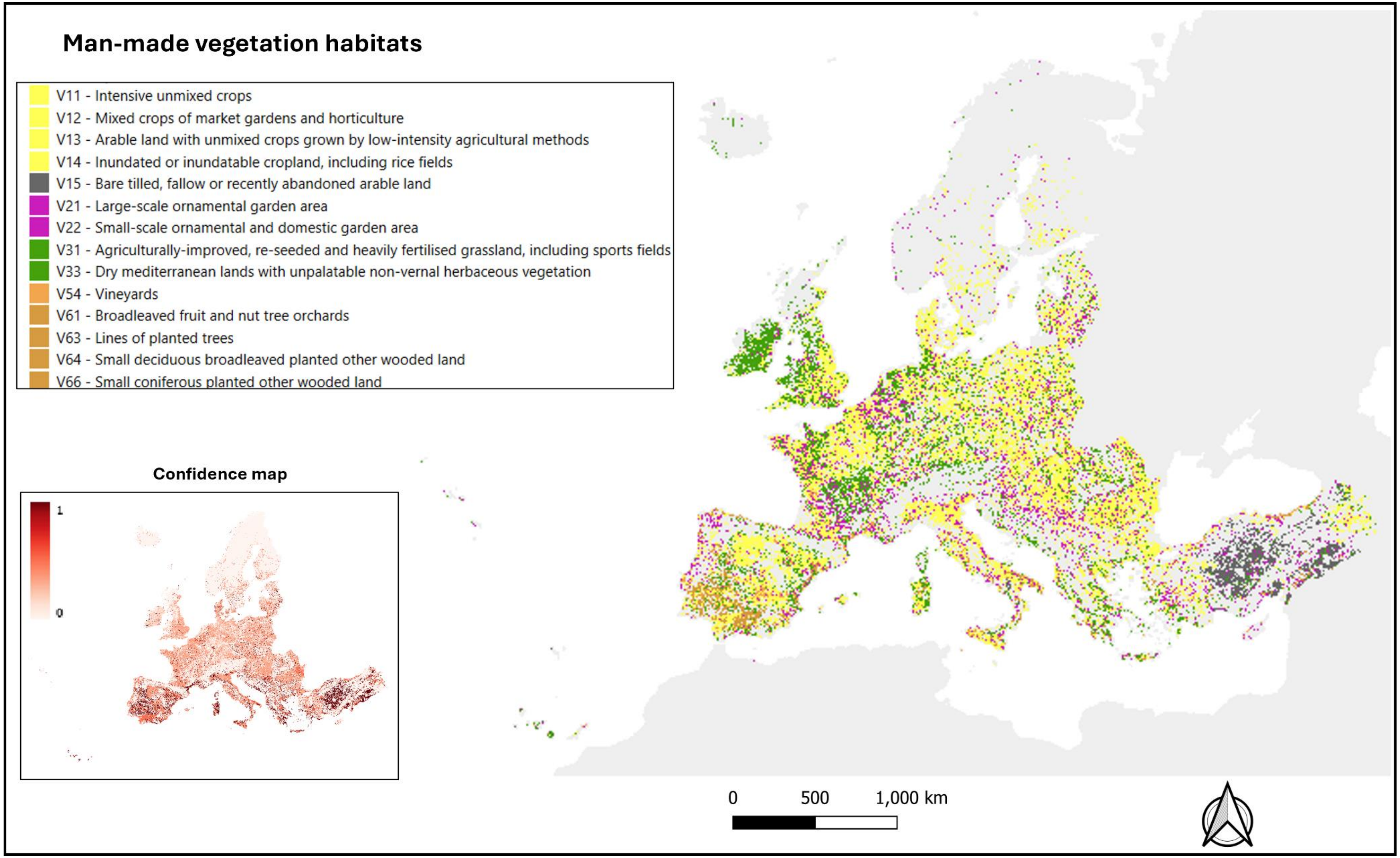

**Figure 10 - Wall-to-wall habitat map - color coded at level 2 for visibility**

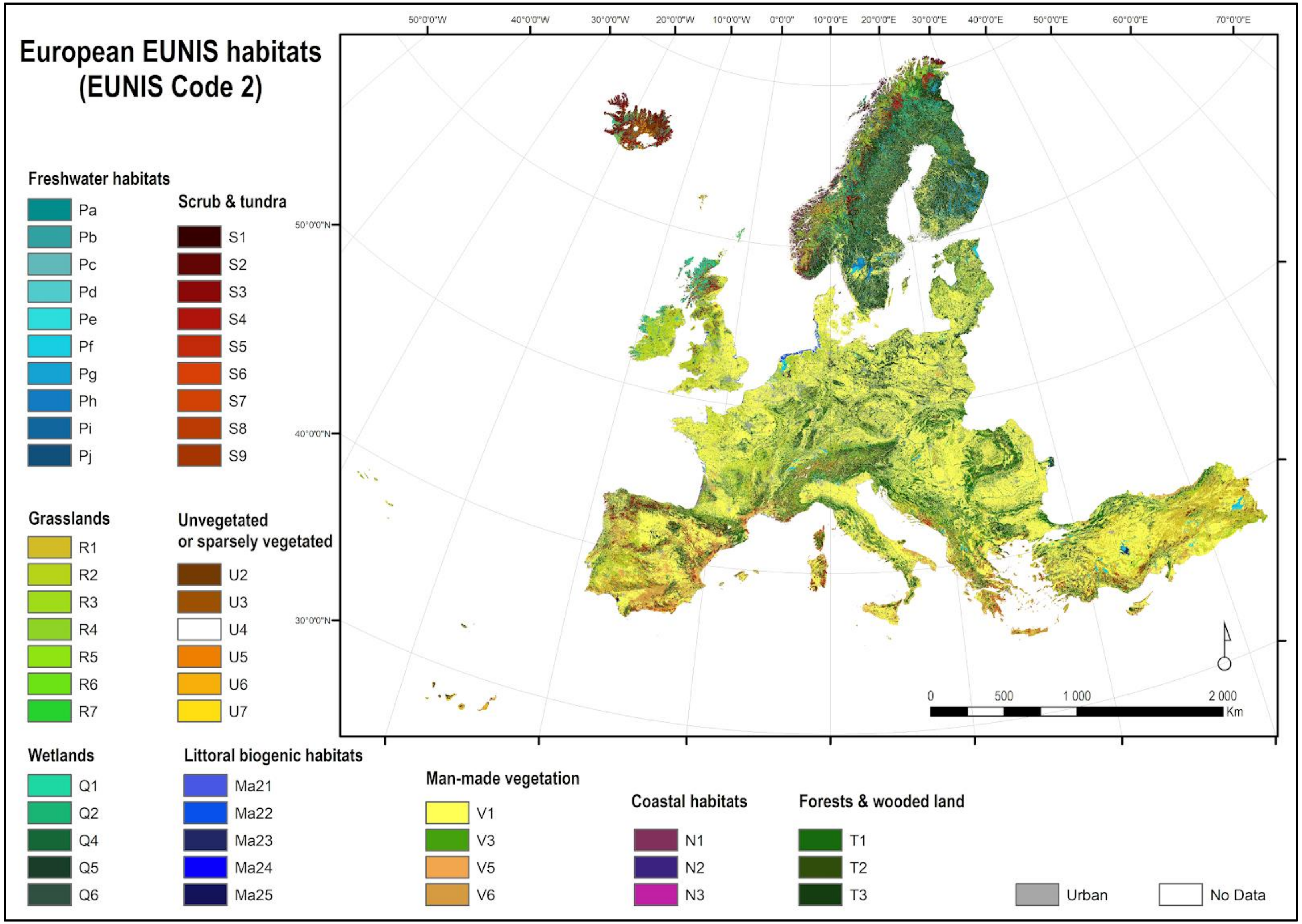